\documentclass{jpsj2}

\title{
Random-Matrix Theory of Electron Transport in
Disordered Wires with Symplectic Symmetry
}

\author{
Hiroshi \textsc{Sakai} and Yositake \textsc{Takane}
}
\inst{%
Department of Quantum Matter, Graduate School of Advanced Sciences of
Matter, Hiroshima University, Higashi-Hiroshima 739-8530
}
\recdate{\today}

\abst{%
The conductance of disordered wires with symplectic symmetry is studied 
by a random-matrix approach. 
It has been believed that Anderson localization inevitably
arises in ordinary disordered wires.
A counterexample is recently found in the systems with symplectic symmetry, 
where one perfectly conducting channel is present 
even in the long-wire limit when the number of conducting channels is odd.
This indicates that the odd-channel case is essentially different from
the ordinary even-channel case. To study such differences, we derive the
DMPK equation for transmission eigenvalues for both the even- and odd-
channel cases. The behavior of dimensionless conductance is investigated
on the basis of the resulting equation.
In the short-wire regime, we find that the weak-antilocalization correction to
the conductance in the odd-channel case is 
equivalent to that in the even-channel case.
We also find that the variance does not depend on whether
the number of channels is even or odd.
In the long-wire regime, 
it is shown that the dimensionless conductance in the even-channel case
decays exponentially as $\langle g_{\mathrm{even}} \rangle \to0$
with increasing system length, while $\langle g_{\mathrm{odd}} \rangle \to1$ 
in the odd-channel case.
We evaluate the decay length for the even- and
odd-channel cases and find a clear even-odd difference.
These results indicate that the perfectly conducting channel induces 
clear even-odd differences in the long-wire regime.
}

\kword{%
disordered quantum wire, random-matrix theory,
symplectic class, conductance
}

\begin{document}
\maketitle

\section{Introduction} 
The statistical aspect of phase-coherent electron transport in quasi-one-dimensional disordered
wires is characterized by three universality classes~\cite{Bee}.
The universality classes describe transport properties which are
independent of the microscopic details of disordered wires.
The orthogonal class consists of systems having both time-reversal symmetry and spin-rotation
invariance, while the unitary class is characterized by the absence of time-reversal symmetry. 
The systems having time-reversal symmetry without spin-rotation invariance
belong to the symplectic class.
In the metallic regime where wire length $L$ is much
shorter than the localization length $\xi$, the weak-localization (weak-antilocalization) effect arises in the
orthogonal (symplectic) class, while the effect disappears in the
unitary class. If the system length $L$ satisfies
$L\gg \xi$, the conductance decays exponentially with
increasing $L$ in all the classes. 
The mean-free path dependence of $\xi$ is different from class to class.
We can systematically analyze such differences by using two different nonperturbative approaches. 
The first is the scaling approach of 
Dorokhov~\cite{Dorok}, and Mello, Pereyra and Kumar~\cite{MPK}, which is based on random-matrix
theory~\cite{meta}.
The second is the super-Fourier analysis by Zirnbauer~\cite{Super1,Super2}, which is 
based on supersymmetric field (SSF) theory of Efetov~\cite{Efetov}.

This work is concerned with the scaling approach.
In terms of the Landauer formula,
the dimensionless conductance $g$ of a disordered wire
is given by the sum of transmission eigenvalues
$\{ T_{a} \}$ of individual channel $a$ as 
\begin{align}
g=\sum_{a=1}^{N} T_{a} ,\label{eq:g}
\end{align}
where $N$ is the number of conducting channels.
The statistical property of the conductance is determined by the
probability distribution function $P(T_{1} ,T_{2} ,...,T_{N} ;L)$
for the transmission eigenvalues.
The scaling approach is based on
an evolution equation of $P(T_{1} ,T_{2} ,...,T_{N} ;L)$
with increasing $L$, which is called the Dorokov-Mello-Pereyra-Kumar (DMPK) equation~\cite{Bee,Dorok,MPK}.
This equation was derived by Dorokhov for the unitary class, and by Mello,
Pereyra and Kumar for the orthogonal class.
Their derivation of the DMPK equation rests on the isotropy assumption
of the distribution of scattering matrices. It is possible
to replace the isotropy assumption by the weaker assumption of
equivalent scattering channels~\cite{MaTom}.
Mac$\hat{\mathrm{e}}$do and Chalker~\cite{MaCha} extended the DMPK equation to the symplectic class
based on the isotropy assumption.

Recent studies show that the transport property of the symplectic class
is very unique compared with the other universality classes~\cite{AN,NA,AS,YW,YT,YT1,YT2,YT3,metallic,M-Caselle}.
From the scattering symmetry ${}^tr=-r$ generally holds in the
symplectic class ($r$: reflection matrix),
we can show that there exists one channel which is
perfectly transmitting without backscattering 
when the number of conducting channels is odd.
The presence of a perfectly conducting channel is first pointed out in
the study of carbon nanotubes~\cite{AN,NA,AS}.
Note that the perfectly conducting channel is present even in the 
limit of $L\to\infty$.
This indicates the absence of Anderson localization in the
odd-channel case. 
Since the odd-channel case is very different from the ordinary
even-channel case, we must separately consider the two cases.
That is, the symplectic class must be divided into the two
subclasses: the symplectic class with an even number of channels and that
with an odd number of channels~\cite{Gruzberg}.
Let $N_{\alpha}$ ($\alpha =\mathrm{even} \ \mathrm{or} \ \mathrm{odd}$)
be the number of conducting channels. 
We consider the dimensionless conductance $g_{\rm even}$ in the
even-channel case with $N_{\rm even} =2m$ and $g_{\rm odd}$ in the
odd-channel case with $N_{\rm odd} =2m+1$, where $m$ is an integer.
In the even-channel case, the number of independent transmission
eigenvalues is $m$ due to Kramers degeneracy, so the dimensionless
conductance is expressed as $g_{\rm even} =2\sum_{a=1}^{m} T_{a}$.
In the odd-channel case, one eigenvalue corresponding to the perfectly
conducting channel is equal to unity, say $T_{2m+1} \equiv 1$,
and other eigenvalues are doubly degenerate. 
Thus, we can express as $g_{\rm odd} =1+2\sum_{a=1}^{m} T_{a}$.
It has been shown that $g_{\rm even} \to 0$ in the long-wire limit of
$L\gg \xi$. In the odd-channel case, we expect 
\begin{align}
g_{\rm odd} \to 1 ,\label{eq:g1}
\end{align}
due to the presence of the perfectly conducting channel.
Such a dramatic even-odd difference does not appear in the orthgonal and
unitary classes.

In this paper, we study the transport property of the 
symplectic class on the basis of the scaling approach.
Our attention is focused on differences between the even- and odd-channel cases. 
We derive the DMPK equation for both the even- and odd-channel cases 
based on the assumption of equivalent scattering channels~\cite{MaTom}.
In the even-channel case, we reproduce the result by
Mac$\hat{\mathrm{e}}$do and Chalker~\cite{MaCha}.
Our assumption used in deriving the DMPK equation 
is weaker than the isotropy assumption of Mac$\hat{\mathrm{e}}$do and Chalker.
The DMPK equation for the odd-channel case is different from that for
the even-channel case reflecting the presence of the perfectly
conducting channel. We study the behavior of the dimensionless
conductance on the basis of the resulting DMPK equation.
The behavior qualitatively changes whether system length $L$ is longer
or shorter than conductance decay length $\xi_{\alpha}$ 
($\alpha =\mathrm{even} \ \mathrm{or} \ \mathrm{odd}$), where $\xi_{\rm even}$
is equivalent to the ordinary localization length.  
Thus, we separately consider the short-wire regime of $L \ll \xi_{\alpha}$
and the long-wire regime of $L \gg \xi_{\alpha}$, but treat the
even- and odd-channel cases in a unified manner.

In the short-wire regime, we obtain the ensemble average and variance of
the dimensionless conductance
using the perturbative treatment by Mello
and Stone~\cite{MaS}.
It is shown that the weak-antilocalization correction to $g_{\rm odd}$
is equivalent to that to $g_{\rm even}$.
Furthermore, we show that the difference between $g_{\rm odd}$ and
$g_{\rm even}$ is very small.
This means that although one additional channel is perfectly conducting
in the odd-channel case, the even-odd difference does not clearly appear
in the short-wire regime.
The reason for this is explained from the viewpoint of the repulsion
between transmission eigenvalues. We show that the repulsion from
the perfectly conducting eigenvalue ($T_{2m+1} \equiv 1$)
to other eigenvalues ($T_{1} ,T_{2} ,...,T_{m}$) is essentially important.

In the long-wire regime, we obtain the probability distribution function 
from the DMPK equation by using the treatment by Pichard~\cite{Pich}.
On the basis of the resulting distribution function, we study the
statistical property of the dimensionless conductance in both the even-
and odd-channel cases. 
We show that the averaged $g_{\alpha}$ behaves as 
$\langle g_{\rm even} \rangle \to 0$
and $\langle g_{\rm odd} \rangle \to 1$
with increasing $L$.
The behavior of $g_{\alpha}$ in this regime is characterized by the conductance
decay length $\xi_{\alpha}$.
To obtain $\xi_{\alpha}$, we define $\Gamma_{\alpha}$ as
$\Gamma_{\rm even} =g_{\rm even} /2$ and 
$\Gamma_{\rm odd} =(g_{\rm even} -1)/2$, and evaluate the average of 
$\ln \Gamma_{\alpha}$.
The conductance decay length is obtained by identifying 
$\exp [\langle \ln \Gamma_{\alpha} \rangle ] \equiv \exp [-2L/\xi_{\alpha}]$.
We find that $\xi_{\rm odd}$ for $N_{\rm odd} =2m+1$ is much shorter
than $\xi_{\rm even}$ for $N_{\rm even} =2m$.
This indicates that $g_{\rm odd}$ decays much faster than $g_{\rm even}$
with increasing $L$. 
Similar to the short-wire regime, the reason for this is explained from
the viewpoint of the eigenvalue repulsion. Again, we show that the
repulsion from $T_{2m+1} \equiv 1$ to other eigenvalues is essentially important.

Our results for both the short- and long-wire regimes are in 
qualitative agreement with those of the super-Fourier analysis~\cite{Super2,YT2}.

The present paper is a detailed and enlarged version of refs. 14
and 18. We here comment on major differences between the present paper
and refs. 14 and 18.
The derivation of the DMPK equation for the odd-channel case and the analysis of $g_{\rm odd}$
in the long-wire limit have been briefly described in ref. 14. However,
a detailed comparison between the even- and odd-channel cases has not
been presented there, so the even-odd difference originated from the
perfectly conducting eigenvalue is less clarified than the present paper.
In ref. 18, the average and variance of $g_{\alpha}$ in the short-wire
regime have been evaluated in terms of a perturbation theory with the
expansion parameter $m^{-1}$. The average $\langle g_{\alpha} \rangle$
is obtained up to $O(m^{-1})$ in ref. 18, while the result up
to $O(m^{-2})$ is given in the present paper. Thus, the present analysis
for the short-wire regime goes one step beyond that in ref. 18. Several
typographical errors in ref. 14 are corrected in the present paper.

In the following, we restrict our consideration to the symplectic class.
The outline of the present paper is as follows.
In the next section, we first adapt the scattering approach to our
problem, and then derive the DMPK equation for both the even- and
odd-channel cases in a unified manner.
The influence of the perfectly conducting eigenvalue becomes clear 
through its derivation.
We also derive the scattering symmetry for the reflection matrix,            
from which the presence of the perfectly conducting channel is proved.
In \S 3, we study the statistical properties of dimensionless conductance $g_{\alpha}$
($\alpha =\mathrm{even} \ \mathrm{or} \ \mathrm{odd}$) focusing on
differences between the even- and odd-channel cases.
We separately treat the short-wire regime of $L\ll \xi_{\alpha}$ and the
long-wire regime of $L\gg \xi_{\alpha}$. We obtain the average and
variance of $g_{\alpha}$ in the short-wire regime. 
A detail of the derivation is given in Appendix.
In the long-wire regime, the average of $\ln \Gamma_{\alpha}$ is
obtained to evaluate the conductance decay length. We also obtain the 
average of $g_{\alpha}$ and the variances of $g_{\alpha}$ and $\ln \Gamma_{\alpha}$. 
The influence of the perfectly conducting
eigenvalue is discussed to explain the observed even-odd differences.
In \S 4, we compare our results with those obtained from the SSF approach.
The conclusion is presented in \S 5.

\section{Scattering Approach to Disordered Wires}
\subsection{Scattering and transfer matrices}
\begin{figure}[t]
\begin{center}
\includegraphics[width=6cm]{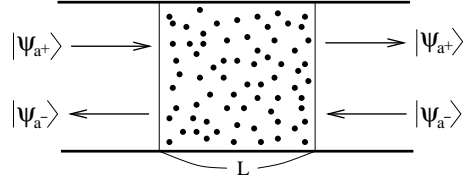}
\end{center}
\caption{Disordered wire of length $L$ sandwiched between two
 ideal leads. In the left lead, $|\psi_{a^{+} } \rangle$ and
 $|\psi_{a^{-} } \rangle$ correspond to the incoming and outgoing
 states, respectively. In the right lead, $|\psi_{a^{+} } \rangle$ and
 $|\psi_{a^{-} } \rangle$ correspond to the outgoing and incoming
 states, respectively.}
\label{f1}
\end{figure}
To apply the scattering approach to our problem, 
we consider a disordered wire sandwiched between ideal leads as
illustrated in Fig.~\ref{f1}. 
Let $N_{\alpha}$ ($\alpha =\mathrm{even} \ \mathrm{or} \ \mathrm{odd}$)
be the number of conducting channels. In this system, electron
transport is described by the scattering matrix $S$, which is a $2N_{\alpha} \times 2N_{\alpha}$
matrix given by
\begin{align}
S=
\left(
\begin{array}{lcr}
r & t^{\prime} \\
t & r^{\prime}
\end{array}
\right)  ,\label{eq:sanran}     %smatrix%
\end{align}
with $N_{\alpha} \times N_{\alpha} $ reflection matrices $r$ and $r^{\prime}$
(reflection from left to left and from right to right) and transmission
matrix $t$ and $t^{\prime}$ (transmission from left to right and from
right to left). 
The Hermitian matrix $tt^{\dagger}$ and
$t^{\prime} t^{\prime \dagger}$ 
have the common set of eigenvalues $T_{1} ,T_{2} ,...,T_{m}$ in the even-channel case of
$N_{\mathrm{even}}=2m$. Each eigenvalue is two-fold degenerate due to Kramers degeneracy.
In the odd-channel case of $N_{\mathrm{odd}}=2m+1$, there arises one additional eigenvalue which
corresponds to the perfectly conducting channel. 
The dimensionless conductance is equal to the sum of all the
transmission eigenvalues, so we obtain
\begin{align}
g_{\mathrm{odd}}  &= 1 + 2\sum_{a=1}^{m} T_{a} ,\label{eq:godd} \\
g_{\mathrm{even}} &= 2 \sum_{a=1}^{m} T_{a}    .\label{eq:geven}    %dimensionless$
\end{align}

We here derive the scattering symmetry for the reflection and
transmission matrices in the symplectic class.
We introduce the Hamiltonian $\mathcal{H}$ 
which is given by
\begin{align}
\mathcal{H} =\mathcal{H}_0 +V ,\label{eq:Hamiltonian}
\end{align} 
where $\mathcal{H}_0$ describes the electronic states 
in the absence of impurities and $V$ is the impurity potential in the
disordered region. 
The above Hamiltonian $\mathcal{H}$ 
has time-reversal symmetry without
spin-rotation invariance.
This means that the ensemble of $\mathcal{H}$ is invariant under 
the transformation $\mathcal{H} \to U\mathcal{H} U^{-1}$, where $U$ is
arbitrary symplectic matrix.
In the absence of disorders (i.e., $V\equiv 0$), we introduce the
propagating wave $|\psi_{a^{+} } \rangle$ in the channel $a$ ($a=1,2,...,N_{\alpha}$)
from left to right (see Fig.~\ref{f1}). 
Note that $|\psi_{a^{+} } \rangle$ is an
eigenfunction of $H_{0}$. We introduce the time-reversal operator 
$U_t$ given by
\begin{align}
U_t = K_0 C ,\label{eq:timeop}
\end{align}
where $C$ is the complex-conjugate operator and 
$K_{0}$ is defined as
\begin{align}
K_0 \equiv
-i\sigma_{y}
.\label{eq:K0}
\end{align}
In terms of the time-reversal operator, we define 
\begin{align}
|\psi_{a^{-}} \rangle 
\equiv
U_{t}
|\psi_{a^{+}} \rangle \label{eq:time-relate}
\end{align}
which represents the propagating wave
in the channel $a$ from right to left. 
Note that $|\psi_{a^{-} } \rangle$ is also an eigenfunction of $H_{0}$.
Using eq.~(\ref{eq:time-relate}) and 
the relation of $U_{t}^2 =-1$,
the eigenfunction $|\psi_{a^{+} } \rangle$ is given by 
\begin{align}
|\psi_{a^{+}} \rangle 
=
-U_{t}
|\psi_{a^{-}} \rangle .\label{eq:time-relate2}
\end{align}
Since our system belongs to the symplectic class,
$\mathcal{H}$ has the time reversal-symmetry as 
\begin{align}
U_t \mathcal{H} U_{t}^{-1} =\mathcal{H}. \label{eq:time-rev}
\end{align}
To observe the symmetry of $r$, we define
\begin{align}
T_{b^{-} a^{+}}
\equiv
\langle \psi_{b^{-} } |
T^{R} |\psi_{a^{+} } \rangle 
,\label{eq:rev-T} 
\end{align}
where $T^{R}$ is the $T$ matrix given by
\begin{align}
T^{R}
\equiv
V
+V\frac{1}{E-H_0 + i\epsilon } V
+V\frac{1}{E-H_0 + i\epsilon } V\frac{1}{E-H_0 + i\epsilon } V
+\cdots \label{eq:Tmatrices}
\end{align}
with $E$ the energy of the scattering state. 
From eqs.~(\ref{eq:timeop}), (\ref{eq:K0}), (\ref{eq:time-rev}) 
and (\ref{eq:Tmatrices}), 
we find that
\begin{align}
U_t T^{R} U_{t}^{-1} 
=
\left(
T^{R} 
\right)^{\dagger}
.\label{eq:Tmat-calcu2}
\end{align}
Using 
eqs.~(\ref{eq:time-relate}), (\ref{eq:time-relate2})
and (\ref{eq:Tmat-calcu2}), 
the  right-hand side of eq.~(\ref{eq:rev-T}) is rewritten as 
\begin{align}
\langle \psi_{b^{-} } |
T^{R} |\psi_{a^{+} } \rangle 
=-\langle U_t \psi_{b^{+}} |
T^{R} |U_t \psi_{a^{-} }
\rangle
=-
\langle
\psi_{a^{-}} |
T^{R}
|\psi_{b^{+}} \rangle
.\label{eq:Tmat-calcu}
\end{align}
Thus, from eqs.~(\ref{eq:rev-T}) and (\ref{eq:Tmat-calcu}), we obtain
$T_{b^{-} a^{+}} =-T_{a^{-} b^{+}}$. This immediately gives the relation
\begin{align}
r_{ba} =
-r_{ab} .\label{eq:Tmat-calcu3}
\end{align}
To observe the symmetry of $t$ and $t^{\prime}$, we define
\begin{align}
T_{b^{-} a^{-}}
\equiv
\langle \psi_{b^{-} } |
T^{R} |\psi_{a^{-} } \rangle 
.\label{eq:tran-T1} 
\end{align}
From eqs.~(\ref{eq:time-relate}), (\ref{eq:time-relate2})
and (\ref{eq:Tmat-calcu2}), the right-hand side of eq.~(\ref{eq:tran-T1})
is rewritten as
\begin{align}
\langle \psi_{b^{-} } |
T^{R} |\psi_{a^{-} } \rangle 
=
\langle U_{t} \psi_{b^{+}} |T^{R} 
| U_{t} \psi_{a^{+}} \rangle
=
\langle \psi_{a^{+}} |T^{R}
|\psi_{b^{+}} \rangle
.\label{eq:tran-T2} 
\end{align}
Thus, from eqs.~(\ref{eq:tran-T1}) and (\ref{eq:tran-T2}), 
we obtain $T_{b^{-} a^{-}}=T_{a^{+} b^{+}}$.
This immediately gives the relation
\begin{align}
t_{ba}^{\prime} 
=
t_{ab}
.\label{eq:tba-tab}
\end{align}
We observe that the transmission and reflection matrices
satisfy~\cite{AS}
\begin{eqnarray}
{}^tt &=& t^{\prime } , \label{eq:timer}  \\
{}^tr &=& -r \ \ \mathrm{and} \ \ 
{}^tr^{\prime }  = -r^{\prime} \label{eq:noback}.
\end{eqnarray}
Equation (\ref{eq:timer}) holds for arbitrary systems with the
time-reversal symmetry if an appropriate set of basisfunctions is chosen, 
while eq.~(\ref{eq:noback}) characterizes the
peculiarity of the symplectic class.
We can show with eq.~(\ref{eq:noback}) that there exists one channel which is
perfectly transmitting without backscattering in the odd-channel case.~\cite{AN,AS}

To study the evolution of transmission eigenvalues, it is
convenient to introduce the transfer matrix $M$ 
expressed by 
\begin{align}
M=
\left(
\begin{array}{lcr}
(t^{\dagger })^{-1}  & r^{\prime} t^{\prime -1} \\
-t^{\prime -1} r     & t^{\prime -1 }   
\end{array}
\right) .\label{eq:transM}                     %Transfer matrix%   
\end{align}
We employ the parameterization~\cite{YW,MaTom}
\begin{align}
M=
\left(
\begin{array}{lcr}
\mathrm{e}^{\theta }  & 0 \\
  0                   & \mathrm{e}^{\theta^{*} }  
\end{array}
\right)
\left(
\begin{array}{ccr}
(1+\eta \eta^{\dagger })^{\frac{1}{2} }  & \eta \\
  \eta^{\dagger }   &   (1+\eta^{\dagger } \eta )^{\frac{1}{2} }
\end{array}
\right)  \label{eq:transM2}
\end{align}
with $\theta =ih$, where $h$ is an arbitrary $N_{\alpha} \times N_{\alpha} $ 
Hermitian matrix and $\eta$ is an arbitrary $N_{\alpha} \times
N_{\alpha}$ complex matrix. They satisfy 
$\theta^{\dagger } =-\theta $ and ${}^{t}\eta = -\eta$. 
Equation (\ref{eq:transM2}) ensures both the flux conservation and the
unique scattering symmetry eqs.~(\ref{eq:timer}) and (\ref{eq:noback}). 
In terms of $\theta$ and $\eta$, we can express the 
transmission and reflection matrices as
\begin{align}
t &= \mathrm{e}^{\theta } (1+\eta \eta^{\dagger } )^{-\frac{1}{2} } ,\label{eq:para1} \\
t^{\prime } &= (1+\eta^{\dagger } \eta )^{-\frac{1}{2} } \mathrm{e}^{-\theta^{*} } ,\label{eq:para2} \\
r &= -(1+\eta^{\dagger } \eta )^{-\frac{1}{2} } \eta^{\dagger }  ,\label{eq:para3} \\
r^{\prime} &= \mathrm{e}^{\theta } \eta (1+\eta^{\dagger } \eta )^{-\frac{1}{2} }
\mathrm{e}^{-\theta^{*} } .  \label{eq:para4} 
\end{align}
We introduce a unitary matrix $v$, reducing $t^{\prime} t^{\prime \dagger}$
to the diagonal form as 
\begin{align} 
v t^{\prime} t^{\prime \dagger} v^{\dagger}
=\mathrm{diag} (T_{1} ,T_{2} ,...,T_{N_{\alpha} } ) .\label{eq:dig}
\end{align}
From eq.~(\ref{eq:para2}), we derive $t^{\prime} t^{\prime \dagger}
=(1+\eta^{\dagger} \eta )^{-1}$ and find that
\begin{align}
v t^{\prime} t^{\prime \dagger} v^{\dagger } =
(1+\hat{\eta}^{\dagger} \hat{\eta} )^{-1} ,\label{eq:dig2}
\end{align}
where $\hat{\eta} =v^{*} \eta v^{\dagger}$. Note that
$\eta^{\dagger} \eta$ has the eigenvalues $\lambda_{1} ,\lambda_{2},...,\lambda_{N_{\alpha}}$
which are given by 
\begin{align}
\lambda_a \equiv 
\frac{1-T_{a} }{T_a } . \label{eq:eigen}
\end{align}
Since $\hat{\eta}$ is the $N_{\alpha} \times N_{\alpha}$
skew-symmetric matrix with ${}^t\hat{\eta} =-\hat{\eta}$,
we obtain
\begin{align}
\hat{\eta}
=
\left(
\begin{array}{@{\,}ccc@{\,}}
0    &  \xi    \\ 
-\xi &  0      \\
\end{array}
\right) \label{eq:etae}
\end{align}
for the even-channel case with $N_{\mathrm{even}} =2m$,
where $\xi$ is an $m\times m$ diagonal matrix given by  
\begin{align}
\xi= \mathrm{diag} \left( \sqrt{\lambda_1} ,\sqrt{\lambda_2} ,...,\sqrt{\lambda_m} \right)
.\label{eq:xi1}
\end{align}
In the odd-channel case with $N_{\mathrm{odd}} =2m+1$, 
we assume without the loss of generality
that the $(2m+1)\mathrm{th}$ channel corresponds to the 
perfectly conducting channel (i.e., $\lambda_{2m+1} =0$). 
This means $T_{2m+1} =1$. We obtain
\begin{align}
\hat{\eta}
=
\left(
\begin{array}{@{\,}ccc@{\,}}
0    & \xi    & 0  \\
-\xi &  0     & \vdots  \\
0    & \ldots & 0               
\end{array}
\right) .\label{eq:etao}
\end{align}
We see that
\begin{align}
\hat{\eta}^{\dagger } \hat{\eta}
=
\mathrm{diag}
\left( 
\lambda_{1} ,\lambda_{2} ,...,
\lambda_{m} ,\lambda_{1} ,\lambda_{2} ,...,
\lambda_{m} 
\right) \label{eq:dig3} 
\end{align}
for the even-channel case and
\begin{align}
\hat{\eta}^{\dagger } \hat{\eta}
=
\mathrm{diag}
\left( 
\lambda_{1} ,\lambda_{2} ,...,
\lambda_{m} ,\lambda_{1} ,\lambda_{2} ,...,
\lambda_{m} ,0
\right) \label{eq:dig4}
\end{align}
for the odd-channel case. We shall use the notation,
$\bar{a} \equiv a+m$ for $1\leq a\leq m$ and $\bar{a} \equiv a-m$ for
$m+1 \leq a \leq 2m$. Using $\bar{a}$, we can rewrite the degeneracy
relation $T_{a} =T_{a+m}$ ($1\leq a\leq m$) as $T_{a} =T_{\bar{a}}$ 
($1\leq a\leq 2m$).

\subsection{DMPK equation}
The distribution of
the transmission eigenvalues 
depends on the wire length $L$. 
Here, we attach the small segment of $\delta x$
in the right-hand side of the system in Fig.~\ref{f1}
and discuss the resulting
deviation of the transmission eigenvalues.
Let $t_{1}^{\prime}$ be the transmission matrix 
$t^{\prime}$ with the wire length $L$ and $t_{2}^{\prime}$
be that with the wire length $L+\delta x$.
We express the transmission eigenvalues of the 
Hermitian matrices $t_{1}^{\prime} t_{1}^{\prime \dagger}$ 
and $t_{2}^{\prime} t_{2}^{\prime \dagger}$ as
$\{ T_{a}\}$ and $\{ \tilde{T_{a}} \}$, respectively.
The transmission eigenvalues $T_{a}$ and
$\tilde{T_{a}}$ are related as
\begin{align}
\tilde{T_{a}} =T_{a} + \delta T_{a} ,\label{eq:deltaTa}   %deviation of transmission eigenvalue%
\end{align}
where $\delta T_{a}$ expresses a deviation. 
We focus on the eigenvalues
from $T_{1}$ to $T_{m}$ and their distribution which depends on $L$.
It has been shown that the evolution of the probability
distribution function $P(T_{1},T_{2},...,T_{m};L)$
with increasing $L$ is generally described by a
Fokker-Planck equation, which is usually called
Dorokhov-Mello-Pereya-Kumar (DMPK) equation in context of the scaling
theory~\cite{Bee}. In terms of $\langle \delta T_{a}
\rangle_{ss}$ and $\langle \delta T_{a} \delta T_{b} \rangle_{ss}$, where 
$\langle \cdots \rangle_{ss}$ denotes the ensemble average with
respect to the small segment of length $\delta x$, 
the DMPK equation is given by~\cite{Bee}
\begin{align}
\frac{\partial P}{\partial L} =
\delta x^{-1}
\sum_{a=1}^{m} \frac{\partial }{\partial T_{a} }
\left( 
-\langle \delta T_{a} \rangle_{ss} P
+\frac{1}{2} \sum_{b=1}^{m}
\frac{\partial}{\partial T_{b} }
\langle \delta T_{a} \delta T_{b} \rangle_{ss} P
\right) .\label{eq:DMPK1}                         %first DMPK%
\end{align}
To obtain an explicit form of the DMPK equation, we must calculate the
ensemble averages $\langle \delta T_{a} \rangle_{ss}$ and
$\langle \delta T_{a} \delta T_{b} \rangle_{ss}$.
To second order in peturbation theory, we obtain
\begin{align}
\delta T_{a} = w_{aa} + \sum_{b(\neq a)}^{N_{\alpha} }
\frac{w_{ab} w_{ba} }{T_{a} -T_{b} } ,\label{eq:setudou}   %puturbation of transmission eigenvalue%
\end{align}
where $w_{ab}$ is an element of the Hermitian matrix 
\begin{align}
w=
t_{2}^{\prime} t_{2}^{\prime \dagger} 
- t_{1}^{\prime} t_{1}^{\prime \dagger} 
\label{eq:t2-t1}
\end{align}
in the basis where $t_{1}^{\prime} t_{1}^{\prime \dagger}$
is diagonal. To relate $t_{1}^{\prime}$ and $t_{2}^{\prime}$,
we introduce the transfer matrix $M_{1}$ for the wire of length $L$
and the transfer matrix $\delta M$ for the small segment. 
The transfer matrix $M_{2}$ for the combined
system is given by
\begin{align}
M_{2} = \delta M M_{1} .\label{eq:combine}   %combine system
\end{align}
We apply the parameterization given in eq.~(\ref{eq:transM2}) 
to $\delta M$. We shall take the weak-scattering limit,~\cite{MaTom} where a moment
higher than the second plays no role, so that $\delta M$ can be
expressed in the simple form 
\begin{align} 
\delta M=
\left(
\begin{array}{lcr}
1+\theta  &  \eta \\
\eta^{\dagger}  &  1+\theta^{*} 
\end{array}
\right) .\label{eq:deltaM}         %delta transfer matrix%
\end{align}
From eqs.~(\ref{eq:transM}), (\ref{eq:combine}) and
(\ref{eq:deltaM}), we can relate $t_{2}^{\prime}$ and $t_{1}^{\prime}$ as 
\begin{align}
t_{2}^{\prime}
=t_{1}^{\prime} \left(
1-\theta^{*} -\eta^{\dagger} r_{1}^{\prime} +\theta^{*2}
\right) . \label{eq:t1t2}
\end{align}
Following Mello and Tomsovic,~\cite{MaTom} 
we assume that
\begin{align}
&\ \langle \theta \rangle_{ss} 
=\langle \eta \rangle_{ss} =0 ,\label{eq:weak01} \\
&\ \langle \theta_{ab} \eta_{cd} \rangle_{ss} 
=\langle \theta_{ab} \eta_{cd}^{\dagger }
\rangle_{ss} =\langle \eta_{ab} \eta_{cd} 
\rangle_{ss} =0                ,\label{eq:weak02} \\
&\ \langle \theta^{2} +
\eta \eta^{\dagger } \rangle_{ss} 
=0                             ,\label{eq:weak03} \\
&\ \langle \theta_{ab} \theta_{cd} \rangle_{ss}
= \kappa_{ab,cd}^{1} ,\label{eq:weak1} \\
&\ \langle \theta_{ab} \theta_{cd}^{*} \rangle_{ss}
= \kappa_{ab,cd}^{2} ,\label{eq:weak2} \\
&\ \langle \eta_{ab} \eta_{cd}^{\dagger} \rangle_{ss}    %weak model%
= \kappa_{ab,cd}^{3} .\label{eq:weak3}
\end{align}
From eqs.~(\ref{eq:t2-t1}) and (\ref{eq:t1t2}),
we decompose the Hermitian matrix $w$ as
\begin{align}
w=w^{(1)} +w^{(2)} ,\label{eq:w1+w2}
\end{align}
where
\begin{align}
w^{(1)}
&=
vt_{1}^{\prime}
\left(
\theta^{*2} +\eta^{\dagger} r_{1}^{\prime}
r_{1}^{\prime \dagger} \eta
\right)t_{1}^{\prime \dagger} v^{\dagger} ,\label{eq:w1} \\
w^{(2)}
&=
vt_{1}^{\prime}
\left(
-r_{1}^{\prime \dagger} \eta
-\eta^{\dagger} r_{1}^{\prime}
\right)
t_{1}^{\prime \dagger} v^{\dagger} .\label{eq:w2}
\end{align}
We find that
\begin{align}
\langle w \rangle_{ss}
&=
\langle w^{(1)} \rangle_{ss} ,\label{eq:ans-w} \\
\langle w w \rangle_{ss}
&= \langle w^{(2)} w^{(2)} \rangle_{ss}  .\label{eq:ans-ww} 
\end{align}
We here take the weak-scattering limit;~\cite{MaTom}
$\delta x$ approaches $0$ and at the same time, we let scattering in the
small segment become infinitely weak, so that
\begin{align}
\sigma_{ab,cd}^{i} 
=\lim_{ \substack{\delta x \to 0 \\ \kappa \to 0} } 
(\delta x)^{-1} \kappa_{ab,cd}^{i} \label{eq:yugen}    
\end{align}
are finite quantities, while $(\delta x)^{-1}$ 
times a moment higher than the second vanishes.
Accordingly, we calculate
\begin{align}
f_{a} 
&=
\lim_{ \substack{\delta x \to 0 \\ \kappa \to 0} } 
(\delta x)^{-1} 
\langle \delta T_{a} \rangle_{ss} ,\label{eq:fa-define} \\
f_{ab} 
&=
\lim_{ \substack{\delta x \to 0 \\ \kappa \to 0} } 
(\delta x)^{-1} 
\langle \delta T_{a} \delta T_{b} \rangle_{ss} ,\label{eq:fab-define} 
\end{align}
instead of $\langle \delta T_{a} \rangle_{ss}$ and 
$\langle \delta T_{a} \delta T_{b} \rangle_{ss}$.
From eqs.~(\ref{eq:setudou}), (\ref{eq:ans-w}) and (\ref{eq:ans-ww}),
we obtain
\begin{align}
f_{a} 
&=
\lim_{ \substack{\delta x \to 0 \\ \kappa \to 0} } 
(\delta x)^{-1} 
\langle w_{aa}^{(1)} \rangle_{ss}
+\lim_{ \substack{\delta x \to 0 \\ \kappa \to 0} } 
(\delta x)^{-1} 
\sum_{b(\neq a)}^{N_{\alpha } }
\frac{1}{T_{a} -T_{b} }
\langle w_{ab}^{(2)} w_{ba}^{(2)} \rangle_{ss} ,\label{eq:fa} \\
f_{ab} 
&=
\lim_{ \substack{\delta x \to 0 \\ \kappa \to 0} } 
(\delta x)^{-1} 
\langle w_{aa}^{(2)} w_{bb}^{(2)} \rangle_{ss} .\label{eq:fab}      %fa and fab%
\end{align}
In accordance with $\theta^{\dagger} =-\theta$ and ${}^{t}\eta =
-\eta$, we adopt the model~\cite{YW} in which $\sigma_{ab,cd}^{i}$ is given by 
\begin{align}
\sigma_{ab,cd}^{1}
&=\lim_{ \substack{\delta x \to 0 \\ \kappa \to 0} } 
(\delta x)^{-1} \langle \theta_{ab} \theta_{cd} \rangle_{ss}
=-\delta_{a,d} \delta_{b,c} \sigma_{ab}^{\prime} , \label{eq:sig1} \\
\sigma_{ab,cd}^{2}
&=\lim_{ \substack{\delta x \to 0 \\ \kappa \to 0} } 
(\delta x)^{-1} \langle \theta_{ab} \theta_{cd}^{*} \rangle_{ss}
=\delta_{a,c} \delta_{b,d} \sigma_{ab}^{\prime} , \label{eq:sig2} \\
\sigma_{ab,cd}^{3}
&=\lim_{ \substack{\delta x \to 0 \\ \kappa \to 0} } 
(\delta x)^{-1} \langle \eta_{ab} \eta_{cd}^{*} \rangle_{ss}
=(\delta_{a,d} \delta_{b,c} - \delta_{a,c} \delta_{b,d}) 
\sigma_{ab} . \label{eq:sig3} 
\end{align}
Here, $\sigma_{ab}$ represents the average
reflection coefficient per unit length from $b$ to $a$, while
$\sigma_{ab}^{\prime}$ represents the average transmission coefficient.
They satisfy $\sum_{b} \sigma_{ab} =\sum_{b} \sigma_{ab}^{\prime}$ since 
$\langle \theta^{2} +\eta \eta^{\dagger} \rangle_{ss} =0$.
To proceed, we must assume that $\sigma_{ab}$ is expressed in a simple form. 
For the even- and odd-channel cases, we adopt the simplest 
choice~\cite{YW} 
\begin{align}
\sigma_{ab} 
=\frac{1-\delta_{a,b} }{(N_{\alpha}
-1)l} ,\label{eq:sigma}
\end{align}
where $l$ is the mean free path. 
This choice corresponds to the equivalent channel model~\cite{MaTom},
and ensures the absence of backscattering
(i.e., $\sigma_{aa} =0$)~\cite{AN,NA}
which characterizes the peculiarity of the symplectic class. 
Equation (\ref{eq:sigma})
leads to
\begin{align}
\sum_{b} \sigma_{ab}
=\sum_{b} \sigma_{ab}^{\prime}
=\frac{1}{l}. \label{eq:meanfree}
\end{align}
From eqs.~(\ref{eq:sig1})-(\ref{eq:meanfree}), we obtain 
\begin{align}
\lim_{ \substack{\delta x \to 0 \\ \kappa \to 0} } 
(\delta x)^{-1} \langle w_{aa}^{(1)} \rangle_{ss}
&=
-\frac{1}{l}
(vt_{1}^{\prime} t_{1}^{\prime \dagger} v^{\dagger})_{aa}
+\frac{1}{(N_{\alpha} -1)l}(vt_{1}^{\prime} t_{1}^{\prime \dagger} v^{\dagger})_{aa}
\cdot \sum_{c=1}^{N_{\alpha}} (1-T_{c} ) \nonumber \\
&-\frac{1}{(N_{\alpha} -1)l}
(vt_{1}^{\prime} r_{1}^{\prime \dagger} r_{1}^{\prime} t_{1}^{\prime
 \dagger} v^{\dagger} )_{aa}  ,\label{eq:limwaa0} \\  
\lim_{ \substack{\delta x \to 0 \\ \kappa \to 0} } 
(\delta x)^{-1} \langle w_{ab}^{(2)} w_{ba}^{(2)} \rangle_{ss}
&=
\frac{1}{(N_{\alpha} -1)l}
\{
(vt_{1}^{\prime} r_{1}^{\prime \dagger} r_{1}^{\prime} t_{1}^{\prime
 \dagger} v^{\dagger})_{aa} \cdot (vt_{1}^{\prime} t_{1}^{\prime
 \dagger} v^{\dagger})_{bb} \nonumber \\
&+(vt_{1}^{\prime} t_{1}^{\prime \dagger}
 v^{\dagger})_{aa} \cdot (vt_{1}^{\prime} r_{1}^{\prime \dagger}
 r_{1}^{\prime} t_{1}^{\prime \dagger} v^{\dagger})_{bb}
-(vt_{1}^{\prime} r_{1}^{\prime \dagger} {}^tt_{1}^{\prime}
{}^tv)_{ab} \nonumber \\
&\times
(v^{*} t_{1}^{\prime *} r_{1}^{\prime} t_{1}^{\prime
 \dagger} v^{\dagger})_{ba}
-(vt_{1}^{\prime} r_{1}^{\prime \dagger} {}^tt_{1}^{\prime}
{}^tv)_{ba} \cdot (v^{*} t_{1}^{\prime *} r_{1}^{\prime} t_{1}^{\prime
 \dagger} v^{\dagger})_{ab}
\}
,\label{eq:limwabba0} \\
\lim_{ \substack{\delta x \to 0 \\ \kappa \to 0} } 
(\delta x)^{-1} \langle w_{aa}^{(2)} w_{bb}^{(2)}
\rangle_{ss}
&=
\frac{1}{(N_{\alpha} -1)l}
\{
(vt_{1}^{\prime} r_{1}^{\prime \dagger} r_{1}^{\prime} t_{1}^{\prime
 \dagger} v^{\dagger})_{ab} \cdot (vt_{1}^{\prime} t_{1}^{\prime
 \dagger} v^{\dagger})_{ba} \nonumber \\
&+(vt_{1}^{\prime} t_{1}^{\prime \dagger}
 v^{\dagger})_{ab} \cdot (vt_{1}^{\prime} r_{1}^{\prime \dagger}
 r_{1}^{\prime} t_{1}^{\prime \dagger} v^{\dagger})_{ba}
 -(vt_{1}^{\prime} r_{1}^{\prime \dagger} {}^tt_{1}^{\prime}
{}^tv)_{ab} \nonumber \\
&\times
(v^{*} t_{1}^{\prime *} r_{1}^{\prime} t_{1}^{\prime
 \dagger} v^{\dagger})_{ab}
-(vt_{1}^{\prime} r_{1}^{\prime \dagger} {}^tt_{1}^{\prime}
{}^tv)_{ba} \cdot (v^{*} t_{1}^{\prime *} r_{1}^{\prime} t_{1}^{\prime \dagger} 
v^{\dagger})_{ba}
\} .\label{eq:limwaabb0}
\end{align} 
Equations (\ref{eq:para1})-(\ref{eq:para4}) give  
\begin{align}
vt_{1}^{\prime} t_{1}^{\prime \dagger} v^{\dagger}
&=
(1+\hat{\eta}_{1}^{\dagger} \hat{\eta}_{1} )^{-1} ,\label{eq:diagonal1} \\
vt_{1}^{\prime} r_{1}^{\prime \dagger} r_{1}^{\prime} t_{1}^{\prime
\dagger} v^{\dagger}
&=
(1+\hat{\eta}_{1}^{\dagger} \hat{\eta}_{1} )^{-2}
\hat{\eta}_{1}^{\dagger} \hat{\eta}_{1} ,\label{eq:diagonal2}\\
vt_{1}^{\prime} r_{1}^{\prime \dagger} {}^tt_{1}^{\prime}
{}^tv
&=
(1+\hat{\eta}_{1}^{\dagger} \hat{\eta}_{1} )^{-\frac{3}{2} }
\hat{\eta}_{1}^{\dagger} ,\label{eq:diagonal3} \\
v^{*} t_{1}^{\prime *} r_{1}^{\prime} t_{1}^{\prime
\dagger} v^{\dagger}
&=
\hat{\eta}_{1} (1+\hat{\eta}_{1}^{\dagger} \hat{\eta}_{1}
 )^{-\frac{3}{2} } ,\label{eq:diagonal4}
\end{align}
where $\hat{\eta}_{1} =v^{*} \eta_{1} v^{\dagger}$. 
From eqs.~(\ref{eq:dig3}), (\ref{eq:dig4})
and (\ref{eq:fa})-(\ref{eq:diagonal4}),
we obtain
\begin{align}
f_{a}
&= -\frac{1}{l} T_{a} +\frac{T_{a}}{(N_{\alpha} -1)l}
\left(
1-T_{a}
+
\sum_{ \substack{b=1 \\ b(\neq a,\bar{a} )} }^{N_{\alpha }}
\frac{T_{a} +T_{b} -2T_{a} T_{b} }{T_{a} -T_{b} }
\right) \nonumber \\
&=
-\frac{1}{l} T_{a} +\frac{T_{a}}{(N_{\alpha} -1)l}
\left(
1-T_{a}
-\delta_{\alpha ,\mathrm{odd}}
+2
\sum_{ \substack{b=1 \\ b(\neq a,\bar{a} )} }^{m}
\frac{T_{a} +T_{b} -2T_{a} T_{b} }{T_{a} -T_{b} }
\right)
\label{eq:fa2}     %fa2%
\end{align}
and 
\begin{align}
f_{ab}
&= \frac{1}{(N_{\alpha} -1)l}
(\delta_{a,b} +\delta_{\bar{a} ,b} )2T_{a}^{2}
(1-T_{a}) \label{eq:fab2}        %fab%
\end{align}
for $1\leq a,b \leq 2m$.
For $a=2m+1$ in the odd-channel case, we also obtain 
\begin{align}
f_{2m+1} &= 0  ,\label{eq:f-0} \\
f_{2m+1\ b} &= 0 .\label{eq:fa2m1} 
\end{align}
In eq.~(\ref{eq:fa2}), we have introduced $\delta_{\alpha ,\mathrm{odd}}$
which is defined as $\delta_{\alpha ,\mathrm{odd}} =1$ for $\alpha =\mathrm{odd}$
and $0$ for $\alpha =\mathrm{even}$.
It should be emphasized that $\delta_{\alpha ,\mathrm{odd}}$ indicates
the contribution from the perfectly conducting eigenvalue.
Substituting eqs.~(\ref{eq:fa2}) and (\ref{eq:fab2}) into
eq.~(\ref{eq:DMPK1}), we obtain 
\begin{align}
\frac{\partial P(T_{1},...,T_{m} ;s)}{\partial s}
=l\sum_{a=1}^{m} \frac{\partial}{\partial T_{a}}
\left(
-f_{a} P(T_{1},...,T_{m};s) +\frac{1}{2}
\frac{\partial}{\partial T_{a}} f_{aa}
P(T_{1},...,T_{m};s)
\right) ,\label{eq:DMPK2} %DMPK2%
\end{align}
where $s=L/l$ is the normalized system length.
Changing variables from $T_{a}$ to 
$\lambda_{a} \equiv (1-T_{a})/T_{a}$,
we finally obtain the DMPK equation 
\begin{eqnarray}
\frac{\partial P(\lambda_{1},...,\lambda_{m} ;s)  }{\partial s} =
\frac{1}{N_{\alpha} -1}
\sum_{a=1}^{m} \frac{\partial }{\partial \lambda_{a} }
\left(
\lambda_{a} (1+\lambda_{a} ) J_{\alpha} 
\frac{\partial}{\partial \lambda_{a} }
\left(\frac{P(\lambda_{1},...,\lambda_{m} ;s) }{J_{\alpha} } \right)
\right).\label{eq:DMPK3}
\end{eqnarray}
The even-odd difference is described by $J_{\alpha }$, 
\begin{align}
J_{\mathrm{even} }
&\equiv  
\prod_{b=1}^{m-1} \prod_{a=b+1}^{m} 
|\lambda_{a} -\lambda_{b} |^{4} ,  \label{eq:eJ} \\
J_{\mathrm{odd} }
&\equiv  
\prod_{c=1}^{m}
\lambda_{c}^{2}
\times
\prod_{b=1}^{m-1} \prod_{a=b+1}^{m} 
|\lambda_{a} -\lambda_{b} |^{4} . \label{eq:oJ} 
\end{align}
Note that the factor $\prod_{c=1}^{m} \lambda_{c}^{2}$ in
$J_{\mathrm{odd} }$ represents the eigenvalue repulsion to 
$\lambda_{1} ,\lambda_{2} ,\cdots ,\lambda_{m}$ arising from the perfectly
conducting eigenvalue of $\lambda_{2m+1} =0$.
This eigenvalue repulsion is the origin of all the even-odd difference.

\section{Transport Properties}
To study the transport properties in disordered wires, it is
convenient to introduce $\Gamma_{\alpha}$ 
($\alpha =\mathrm{even} \ \ \mathrm{or} \ \ \mathrm{odd}$)
which is defined as
\begin{align}
\Gamma_{\alpha} \equiv \sum_{a=1}^{m} T_{a}
=\sum_{a=1}^{m} \frac{1}{1+\lambda_{a} } .\label{eq:Gammadef}  %Gamma
\end{align}
From eqs.~(\ref{eq:godd}), (\ref{eq:geven}) and (\ref{eq:Gammadef}),
the dimensionless conductance for the even- and
odd-channel cases is given by
\begin{align} 
g_{\alpha} 
=
\delta_{\alpha ,\mathrm{odd} }
+2 \Gamma_{\alpha} 
.\label{eq:ensemble}                      %ensemble average conductance
\end{align}

\subsection{Short-wire regime} 
We adopt Mello and Stone's method~\cite{MaS}
to obtain the average and variance of $g_{\alpha}$,
\begin{align} 
\langle g_{\alpha} \rangle
&=
\delta_{\alpha ,\mathrm{odd} }
+2\langle \Gamma_{\alpha} \rangle 
,\label{eq:ensemble2} \\                     %ensemble average conductance
\mathrm{Var} [g_{\alpha} ]
&= 4\left(
\langle \Gamma_{\alpha}^{2} \rangle -\langle
\Gamma_{\alpha} \rangle^{2} 
\right) , \label{eq:var1}
\end{align}
where
\begin{eqnarray}
\langle \cdots \rangle
= \int \prod_{a=1}^{m} \mathrm{d} \lambda_{a} \cdots P(\lambda_{1}
,...,\lambda_{m} ;\ s). \label{eq:enave} 
\end{eqnarray}
From eq.~(\ref{eq:DMPK3}), we can derive the evolution equation for the
ensemble average $\langle F(\lambda_{1} ,...,\lambda_{m} ) \rangle$
of an arbitrary function $F(\lambda_{1} ,...,\lambda_{m} )$.~\cite{MaS}
Multiplying both sides of eq.~(\ref{eq:DMPK3}) by $F$ and integrating
over $\{\lambda_{a} \}$, we obtain,
\begin{eqnarray}
(N_{\alpha } -1)\frac{\partial \langle F \rangle }{\partial s}
&=&
\Biggl\langle \sum_{a=1}^{m}
\biggl\{
\lambda_{a} (1+\lambda_{a} )
\frac{\partial^{2} F}{\partial \lambda_{a}^{2} }
+ (1+2 \lambda_{a} )\frac{\partial F}{\partial \lambda_{a} }
\biggr\} \nonumber \\ 
&\ & \ \ +2 \sum_{a=1}^{m} \sum_{\substack{b=1 \\b (\neq a)}}^{m}
\frac{\lambda_{a} (1+ \lambda_{a} ) \frac{\partial F}{\partial \lambda_{a} }   
-\lambda_{b} (1+ \lambda_{b} ) \frac{\partial F}{\partial \lambda_{b} }}{
\lambda_{a}  -\lambda_{b} }  \Biggr\rangle  \nonumber \\
&\ & \ \ \ \ \ \ \ \ \ \ 
+ \delta_{\alpha ,\mathrm{odd}} 
\Biggl\langle 
\sum_{a=1}^{m}
\biggl\{
2(1+\lambda_{a} ) \frac{ \partial F}{ \partial \lambda_{a} } 
\biggr\}
\Biggr\rangle .\label{eq:Ffunc}
\end{eqnarray}
Let us derive the evolution equation for $\langle \Gamma_{\alpha} \rangle$.
Setting $F=\Gamma_{\alpha }$ in eq.~(\ref{eq:Ffunc}), we obtain  
\begin{align}
(N_{\alpha } -1) \frac{\partial \langle \Gamma_{\alpha} \rangle }{\partial s}
=\bigl\langle -2\Gamma_{\alpha }^{2} 
+\Gamma_{\alpha ,2} -2\Gamma_{\alpha }
\delta_{\alpha ,\mathrm{odd} } \bigr\rangle ,\label{eq:Gamma1} 
\end{align}
where $\Gamma_{\alpha q}$ was defined as $\Gamma_{\alpha q} =\sum_{a=1}^{m} T_{a}^{q} $. 
To solve eq.~(\ref{eq:Gamma1}), we need the quantities 
$\langle \Gamma_{\alpha }^{2} \rangle$ and 
$\langle \Gamma_{\alpha 2} \rangle$.
So we set $F=\langle \Gamma_{\alpha 2} \rangle$
in eq.~(\ref{eq:Ffunc}) and obtain
\begin{align}
(N_{\alpha } -1) \frac{\partial \langle \Gamma_{\alpha 2} \rangle }{\partial s}
&=
\bigl\langle 
4\Gamma_{\alpha }^{2} -8\Gamma_{\alpha } \Gamma_{\alpha 2}
-2(1+2\delta_{\alpha ,\mathrm{odd} } )\Gamma_{\alpha 2} 
+4\Gamma_{\alpha 3}
\bigr\rangle .\label{eq:sample2}
\end{align}
New quantities $\langle \Gamma_{\alpha}^{2} \rangle$, 
$\langle \Gamma_{\alpha} \Gamma_{\alpha 2} \rangle$
and $\langle \Gamma_{\alpha 3} \rangle$ 
appear in the right-hand side of eq.~(\ref{eq:sample2}). 
We will observe that every time we write down an evolution equation, we find, on the 
right-hand side, new quantities that have not appeared before. 
This means that we cannot obtain a closed set of coupled equations, and 
thereby cannot obtain an exact solution. To overcome this difficulty, 
we employ the method by Mello and Stone~\cite{MaS}
to obtain an approximate solution for $\langle \Gamma_{\alpha} \rangle$ and
$\langle \Gamma_{\alpha}^{2} \rangle$.
This method is based on a series expansion with respect to $m^{-1}$,
and is applicable to the short-wire regime of $s\ll m$.
Thus, we assume in our following argument that $m$ is much larger than unity. 
We derive the evolution equation for $\langle \Gamma_{\alpha}^{p} \rangle$,
as well as those for the ensemble averages 
$\langle \Gamma_{\alpha}^{p} \Gamma_{\alpha 2}\rangle$, 
$\langle \Gamma_{\alpha}^{p} \Gamma_{\alpha 3} \rangle$,
$\langle \Gamma_{\alpha}^{p} \Gamma_{\alpha 2}^{2} \rangle$,
$\langle \Gamma_{\alpha}^{p} \Gamma_{\alpha 4} \rangle$,
$\langle \Gamma_{\alpha}^{p} \Gamma_{\alpha 2} \Gamma_{\alpha 3}
\rangle$ and 
$\langle \Gamma_{\alpha}^{p} \Gamma_{\alpha 2}^{3} \rangle$.  
Setting 
$F=\Gamma_{\alpha }^{p} ,\ \Gamma_{\alpha}^{p} \Gamma_{\alpha 2},\
\Gamma_{\alpha}^{p} \Gamma_{\alpha 3} ,\ \Gamma_{\alpha}^{p}
\Gamma_{\alpha 2}^{2} ,\ \Gamma_{\alpha}^{p} \Gamma_{\alpha 4} ,\
\Gamma_{\alpha}^{p} \Gamma_{\alpha 2} \Gamma_{\alpha 3} $ and
$\Gamma_{\alpha}^{p} \Gamma_{\alpha 2}^{3}$ in eq.~(\ref{eq:Ffunc}), 
we obtain the evolution equations 
\begin{eqnarray}
(N_{\alpha } -1) \frac{\partial \langle \Gamma_{\alpha}^{p} \rangle }{\partial s}
&=&\bigl\langle -2p \Gamma_{\alpha }^{p+1} +p \Gamma_{\alpha }^{p-1} \Gamma_{\alpha 2} 
\nonumber \\
&\ &
+p(p-1) \Gamma_{\alpha }^{p-2} (\Gamma_{\alpha 2} -\Gamma_{\alpha 3} ) -2p\Gamma_{\alpha }^{p}
\delta_{\alpha ,\mathrm{odd} } \bigr\rangle , \label{eq:Tfunc} \\
(N_{\alpha } -1) \frac{\partial \langle \Gamma_{\alpha}^{p} \Gamma_{\alpha 2}
\rangle }{\partial s}
&=& 
\bigl\langle 4 \Gamma_{\alpha }^{p+2} 
-2(p+4) \Gamma_{\alpha }^{p+1} \Gamma_{\alpha 2}  \nonumber \\
&\ &
-2\{ 1+(p+2)\delta_{\alpha ,\mathrm{odd} }
\} \Gamma_{\alpha }^{p} \Gamma_{\alpha 2} 
+4\Gamma_{\alpha }^{p} \Gamma_{\alpha 3}  \nonumber \\
&\ &
+p\Gamma_{\alpha}^{p-1} \Gamma_{\alpha 2}^{2}
+4p\Gamma_{\alpha}^{p-1} (\Gamma_{\alpha 3} - \Gamma_{\alpha 4} ) \nonumber \\
&\ &
+p(p-1)\Gamma_{\alpha }^{p-2} (\Gamma_{\alpha 2}^{2} -\Gamma_{\alpha 2} \Gamma_{\alpha 3} )
\bigr\rangle , \label{eq:T2func}  \\
(N_{\alpha } -1) \frac{\partial \langle \Gamma_{\alpha}^{p} \Gamma_{\alpha 3}
\rangle }{\partial s}
&=&
\bigl\langle 
-2(p+6)
\Gamma_{\alpha }^{p+1} \Gamma_{\alpha 3}
+12 \Gamma_{\alpha }^{p+1} \Gamma_{\alpha 2}
-6 \Gamma_{\alpha }^{p} \Gamma_{\alpha 2}^{2} \nonumber \\
&\ &
-2\{ 3+(p+3)\delta_{\alpha ,\mathrm{odd} }
\} 
\Gamma_{\alpha }^{p} \Gamma_{\alpha 3} 
+9\Gamma_{\alpha }^{p} \Gamma_{\alpha 4} \nonumber \\
&\ & +p\Gamma_{\alpha}^{p-1} \Gamma_{\alpha 2} \Gamma_{\alpha 3}
+6p\Gamma_{\alpha }^{p-1}
(\Gamma_{\alpha 4} -\Gamma_{\alpha 5} ) \nonumber \\
&\ &
+p(p-1)\Gamma_{\alpha }^{p-2}
(\Gamma_{\alpha 2} \Gamma_{\alpha 3} - \Gamma_{\alpha 3}^{2} )
\bigr\rangle , \label{eq:T3func} \\
(N_{\alpha } -1) \frac{\partial \langle \Gamma_{\alpha}^{p} \Gamma_{\alpha 2}^{2}
\rangle }{\partial s}
&=&
\bigl\langle 
-2(p+8)
\Gamma_{\alpha }^{p+1} \Gamma_{\alpha 2}^{2}
+8 \Gamma_{\alpha }^{p+2} \Gamma_{\alpha 2}
+p \Gamma_{\alpha }^{p-1} \Gamma_{\alpha 2}^{3} \nonumber \\
&\ &
-2\{
2+(p+4) \delta_{\alpha ,\mathrm{odd} } 
\} \Gamma_{\alpha }^{p} \Gamma_{\alpha 2}^{2}
+8\Gamma_{\alpha}^{p} \Gamma_{\alpha 2} \Gamma_{\alpha 3} \nonumber \\ 
&\ &
+p(p-1)\Gamma_{\alpha }^{p-2} (\Gamma_{\alpha 2} - \Gamma_{\alpha 3} ) \Gamma_{\alpha 2}^{2}
+8\Gamma_{\alpha}^{p} (\Gamma_{\alpha 4} -\Gamma_{\alpha 5 } )  \nonumber \\
&\ &
+8p\Gamma_{\alpha }^{p-1} \Gamma_{\alpha 2} (\Gamma_{\alpha 3} -\Gamma_{\alpha 4} )
\bigr\rangle ,\label{eq:T22func} \\
(N_{\alpha } -1) \frac{\partial \langle \Gamma_{\alpha}^{p}
 \Gamma_{\alpha 4} \rangle}{\partial s}
&=&
\bigl\langle
p(p-1)\Gamma_{\alpha}^{p-2} (\Gamma_{\alpha 2} -\Gamma_{\alpha 3}
)\Gamma_{\alpha 4} 
+p\Gamma_{\alpha }^{p-1} \Gamma_{\alpha 2} \Gamma_{\alpha 4} 
\nonumber \\
&\ &
-2(p+8)\Gamma_{\alpha }^{p+1} \Gamma_{\alpha 4}
-2\{ (p+4)\delta_{\alpha ,\mathrm{odd} } +6 \}
\Gamma_{\alpha }^{p} \Gamma_{\alpha 4} 
\nonumber \\
&\ &
+16\Gamma_{\alpha }^{p}
\Gamma_{\alpha 5} 
+8p\Gamma_{\alpha }^{p-1} (\Gamma_{\alpha 5} -\Gamma_{\alpha 6}
)+16\Gamma_{\alpha }^{p+1} \Gamma_{\alpha 3} 
\nonumber \\
&\ &
+8\Gamma_{\alpha }^{p} \Gamma_{\alpha 2}^{2} -16\Gamma_{\alpha }^{p}
\Gamma_{\alpha 2} \Gamma_{\alpha 3}
\bigr\rangle ,\label{eq:TT4} \\
(N_{\alpha } -1) \frac{\partial \langle \Gamma_{\alpha}^{p}
 \Gamma_{\alpha 2} \Gamma_{\alpha 3} \rangle}{\partial s}
&=&
\bigl\langle
p(p-1)\Gamma_{\alpha }^{p-2} \Gamma_{\alpha 2} \Gamma_{\alpha 3}
(\Gamma_{\alpha 2} -\Gamma_{\alpha 3} )
\nonumber \\
&\ &
+4p\Gamma_{\alpha }^{p-1} \Gamma_{\alpha 3} (\Gamma_{\alpha 3}
-\Gamma_{\alpha 4} ) 
+6p\Gamma_{\alpha }^{p-1} \Gamma_{\alpha 2} (\Gamma_{\alpha 4}
-\Gamma_{\alpha 5} )
\nonumber \\
&\ &
+12\Gamma_{\alpha }^{p} (\Gamma_{\alpha 5} -\Gamma_{\alpha 6} )
+4\Gamma_{\alpha }^{p} \Gamma_{\alpha 3}^{2} 
+p\Gamma_{\alpha }^{p-1} \Gamma_{\alpha 2}^{2} \Gamma_{\alpha 3}
\nonumber \\
&\ &
-2
\{ 
4+(p+5)\delta_{\alpha ,\mathrm{odd}}
\}
\Gamma_{\alpha }^{p} \Gamma_{\alpha 2} \Gamma_{\alpha 3}
+9\Gamma_{\alpha }^{p} \Gamma_{\alpha 2} \Gamma_{\alpha 4}
\nonumber \\
&\ &
+4\Gamma_{\alpha }^{p+2} \Gamma_{\alpha 3}
-2(p+10)\Gamma_{\alpha }^{p+1} \Gamma_{\alpha 2} \Gamma_{\alpha 3}
+12\Gamma_{\alpha }^{p+1} \Gamma_{\alpha 2}^{2} 
\nonumber \\
&\ &
-6\Gamma_{\alpha }^{p} \Gamma_{\alpha 2}^{3}
\bigr\rangle ,\label{eq:TT2T3} \\
(N_{\alpha } -1) \frac{ \partial \langle \Gamma_{\alpha }^{p}
 \Gamma_{\alpha 2}^{3} \rangle
 }{\partial s}
&=&
\bigl\langle
p(p-1)\Gamma_{\alpha }^{p-2} (\Gamma_{\alpha 2} -\Gamma_{\alpha 3}
)\Gamma_{\alpha 2}^{3}
+p\Gamma_{\alpha }^{p-1} \Gamma_{\alpha 2}^{4} 
\nonumber \\
&\ &
+24\Gamma_{\alpha }^{p}
\Gamma_{\alpha 2}
(\Gamma_{\alpha 4} -\Gamma_{\alpha 5} )
-2p(p+12)\Gamma_{\alpha}^{p+1} \Gamma_{\alpha 2}^{3}
\nonumber \\
&\ &
+12\Gamma_{\alpha }^{p} \Gamma_{\alpha 2}^{2} \Gamma_{\alpha 3}
+12p\Gamma_{\alpha }^{p-1} \Gamma_{\alpha 2}^{2}
(\Gamma_{\alpha 3} -\Gamma_{\alpha 4} )
\nonumber \\
&\ &
+12\Gamma_{\alpha }^{p+2} \Gamma_{\alpha 2}^{2}
-2
\{
3+(p+6)\delta_{\alpha, \mathrm{odd}} 
\}
\Gamma_{\alpha}^{p} \Gamma_{\alpha ,2}^{3}
\bigr\rangle .\label{eq:TT23}
\end{eqnarray}
We seek the solution for the above equations as a series in decreasing
powers of $m$. To do so, we expand the averages as
\begin{eqnarray} 
\langle \Gamma_{\alpha }^{p} \rangle
&=& 
m^{p} f_{\alpha , p,0} (s) +m^{p-1} f_{\alpha ,p,1} (s)
+m^{p-2} f_{\alpha ,p,2} (s) 
+m^{p-3} f_{\alpha ,p,3} (s)
+ \cdots ,\ \ \ \label{eq:fseri} \\
\langle \Gamma_{\alpha }^{p} \Gamma_{\alpha 2} \rangle
&=& 
m^{p+1} j_{\alpha , p+1,0} (s) +m^{p} j_{\alpha ,p+1,1} (s)
+m^{p-1} j_{\alpha ,p+1,2} (s) + \cdots ,\label{eq:gseri} \\
\langle \Gamma_{\alpha }^{p} \Gamma_{\alpha 3} \rangle
&=& 
m^{p+1} k_{\alpha , p+1,0} (s) +m^{p} k_{\alpha ,p+1,1} (s)
+ \cdots ,\label{eq:hseri} \\
\langle \Gamma_{\alpha }^{p} \Gamma_{\alpha 2}^{2} \rangle
&=& 
m^{p+2} l_{\alpha , p+2,0} (s) +m^{p+1} l_{\alpha ,p+2,1} (s)
+ \cdots ,\label{eq:lseri}  \\
\langle \Gamma_{\alpha }^{p} \Gamma_{\alpha 4} \rangle 
&=&
m^{p+1} t_{\alpha ,p+1,0} (s)+\cdots ,\label{eq:tseri} \\
\langle \Gamma_{\alpha}^{p} \Gamma_{\alpha 2} \Gamma_{\alpha 3} \rangle
&=&
m^{p+2} u_{\alpha ,p+2,0} (s) +\cdots ,\label{eq:useri} \\
\langle \Gamma_{\alpha }^{p} \Gamma_{\alpha 2}^{3} \rangle 
&=&
m^{p+3} y_{\alpha ,p+3,0} (s) +\cdots ,\label{eq:yseri}
\end{eqnarray}
where $f_{\alpha ,p,n} (s)$, $j_{\alpha ,p,n} (s)$, $k_{\alpha ,p,n}
(s)$, $l_{\alpha ,p,n} (s)$, $t_{\alpha ,p,n} (s)$, $u_{\alpha ,p,n} (s)$
and $y_{\alpha ,p,n} (s)$ 
are functions of $s$ and satisfy the initial conditions
\begin{align}
f_{\alpha ,p,n} (0)=j_{\alpha ,p,n} (0)=k_{\alpha ,p,n} (0)=l_{\alpha
,p,n} (0)=t_{\alpha ,p,n} (0)=u_{\alpha ,p,n} (0)=y_{\alpha ,p,n} (0)=\delta_{n,0}
.\label{eq:initial}
\end{align}
From eqs.~(\ref{eq:Tfunc})-(\ref{eq:initial}),
we evaluate $f_{\alpha ,p,0} (s)$, $f_{\alpha ,p,1}
(s)\cdots$. The detail of calculations is given in Appendix.
Substituting eqs.~(\ref{eq:f0}),
(\ref{eq:f1kai}), (\ref{eq:f2kai}) and 
(\ref{eq:fp3s}) into eq.~(\ref{eq:fseri}),
we finally obtain
\begin{align} 
\langle \Gamma_{\alpha }^{p} \rangle
&=
m^{p} \frac{1}{(1+s)^{p} } 
+m^{p-1}
\frac{p}{6(1+s)^{p+2} }
\left[
s^{3} -(3s^{3} +6s^{2} +3s)\delta_{\alpha ,\mathrm{odd} }
\right] \nonumber \\
&+ m^{p-2} \frac{p}{360(1+s)^{p+4}}
\Bigl[
\{ 11p-9+(15p-15)\delta_{\alpha ,\mathrm{odd} } \} s^{6} 
\nonumber \\
& \ \ \ \ \ \ \ \ \ \ \ \ \ \ \ \ \ \ \ \ \ \ \ \ \ \ \ \ \ \ \ \ \ \ 
+
\{ 36p-24+(120p-120)\delta_{\alpha ,\mathrm{odd} } \} s^{5}
\nonumber \\
& \ \ \ \ \ \ \ \ \ \ \ \ \ \ \ \ \ \ \ \ \ \ \ \ \ \ \ \ \ \ \ \ \ \ 
+
\{ 90p-60+(240p-240)\delta_{\alpha ,\mathrm{odd} }\} s^{4}
\nonumber \\
& \ \ \ \ \ \ \ \ \ \ \ \ \ \ \ \ \ \ \ \ \ \ \ \ \ \ \ \ \ \ \ \ \ \ 
+
\{ 120p-150+(180p-180)\delta_{\alpha ,\mathrm{odd} }\} s^{3}
\nonumber \\
& \ \ \ \ \ \ \ \ \ \ \ \ \ \ \ \ \ \ \ \ \ \ \ \ \ \ \ \ \ \ \ \ \ \ 
+
\{ 90p-90+(45p-45)\delta_{\alpha ,\mathrm{odd} }\} s^{2}
\Bigr]
\nonumber \\
&+m^{-3} \frac{p}{45360(1+s)^{p+6} }
\Bigl[
\{
(161p^2 -369p+196)
\nonumber \\
&\ \ \ \ \ \ \ \ \ \ \ \ \ \ \ \ \ \ \ \ \ \ \ \ \ \ \ \ 
-(693p^2 -1953p+1260)\delta_{\alpha ,\mathrm{odd}}
\} s^9
\nonumber \\
&+
\{
(756p^2 -1368p+630)
-(5544p^2 -15624p+10206)\delta_{\alpha ,\mathrm{odd} }
\} s^8
\nonumber \\
&+
\{
(1890p^2 -2448p+1008)
-(19404p^2 -54684p +36288)\delta_{\alpha ,\mathrm{odd} }
\} s^7
\nonumber \\
&+
\{
(2520p^2 -2142p+1260)-(36288p^2 -106218p+73458)\delta_{\alpha ,\mathrm{odd} }
\} s^6 \ \ \ \ \ 
\nonumber \\
&+
\{
(1890p^2 +3402p-4914)-(39690p^2 -120960p+83916)\delta_{\alpha ,\mathrm{odd} }
\} s^5
\nonumber \\
&+
\{
(9450p-24570)-(24570p^2 -75600p+49140)\delta_{\alpha ,\mathrm{odd} }
\} s^4
\nonumber \\
&+
\{
(18900p-24570)-(6615p^2 -19845p+11340)\delta_{\alpha ,\mathrm{odd} }
\} s^3
\nonumber \\
&+
(5670p-5670)s^2
\Bigr] +\cdots .\label{eq:Gamma} %Gamma
\end{align}
Setting $p=1$ in eq.~(\ref{eq:Gamma}) and substituting the resulting
expression for $\langle \Gamma_{\alpha} \rangle$ into
eq.~(\ref{eq:ensemble2}), we obtain
\begin{align}
\langle g_{\mathrm{odd}} \rangle
&=
(2m+1) \frac{1}{1+s} 
+\frac{s^{3} }{3(1+s)^{3} } 
+m^{-1} \frac{1}{90(1+s)^{5} } (s^{6} +6s^{5} +15s^{4} -15s^{3} )
\nonumber \\
&-
m^{-2} \frac{1}{3780(1+s)^7 } (2s^9 +18s^8 +93s^7 +315s^6 +378s^5 +2205s^4
+630s^3 )
\nonumber \\
&\ \ \ \ \ \ \ \ +O \left( m^{-3} \right) 
,\label{eq:go} \\ %godd 
\langle g_{\mathrm{even}} \rangle
&=
2m \frac{1}{1+s} 
+\frac{s^{3} }{3(1+s)^{3} }
+m^{-1} \frac{1}{90(1+s)^{5} } (s^{6} +6s^{5} +15s^{4} -15s^{3} )
\nonumber \\
&-
m^{-2} \frac{1}{3780(1+s)^7 } (2s^9 -3s^8 -75s^7 -273s^6 -63s^5 +2520s^4
+945s^3 )
\nonumber \\
&\ \ \ \ \ \ \ \ +O \left( m^{-3} \right) . \label{eq:ge}    %geven
\end{align}

Now, we discuss the behavior of the averaged conductance in the short-wire
regime of $s\ll m$. The first term in $\langle g_{\alpha } \rangle$     %metallic
represents the classical contribution.
The second term in $\langle g_{\alpha } \rangle$ represents the  
weak-antilocalization correction. We observe that the weak-antilocalization
correction to $\langle g_{\mathrm{odd}} \rangle$ is equivalent to that to            
$\langle g_{\mathrm{even}} \rangle$.
We also observe that the third term in $\langle g_{\mathrm{odd}} \rangle$
is equivalent to that in $\langle g_{\mathrm{even}} \rangle$.
We find a small even-odd difference in the forth terms of $\langle g_{\mathrm{odd}} \rangle$ and $\langle g_{\mathrm{even}} \rangle$, 
although their leading order corrections are equivalent to each other.
This indicates that, in the short-wire regime, a notable 
even-odd difference does not appear in the averaged
conductance. We here discuss the reason why $\langle g_{\mathrm{odd}} \rangle$
is almost equivalent to $\langle g_{\mathrm{even}} \rangle$.
Note that the perfectly conducting
channel is present only in the odd-channel case, and that if we neglect
this special channel, the total number of conducting channels 
in the odd-channel case is equivalent to that in the even-channel case. 
Thus, one may expect 
\begin{align}
\langle g_{\mathrm{odd}}
\rangle \stackrel{?}{=} \langle g_{\mathrm{even}} \rangle  +1 .\label{eq:gyosou}
\end{align}
However, contrary to this hypothesis, our result indicates that
$\langle g_{\mathrm{odd}} \rangle \sim \langle g_{\mathrm{even}} \rangle$.
To explain this, we must consider the repulsion acting between
transmission eigenvalues of $\{ T_{a}\}$.
In the odd-channel case, one transmission
eigenvalue $T_{2m+1}$ is equal to unity due to the presence of the perfectly
conducting channel. Although it positively contributes to 
$\langle g_{\mathrm{odd} } \rangle$, the other eigenvalues
$\{ T_{1},...,T_{m} \}$ are reduced 
due to the repulsion from it. 
To observe this, we set $p=1$ in eq.~(\ref{eq:Gamma}) and present the expression
for $\Gamma_{\alpha}$,
\begin{align}
\langle \Gamma_{\alpha } \rangle
=
m \frac{1}{(1+s)} 
+\frac{s^{3}}{6(1+s)^{3} } 
-\frac{s}{2(1+s)} \delta_{\alpha ,\mathrm{odd} }
+ O\left(  m^{-1} \right) . \label{eq:Gammap1}
\end{align}
The third term in eq.~(\ref{eq:Gammap1})
represents this reduction. Our result indicates that the contribution to 
$\langle g_{\mathrm{odd}} \rangle$ from the perfectly conducting channel
is almost canceled out by the reduction of the contribution from the other channels.

Next, we obtain the variance $\mathrm{Var} [g_{\mathrm{\alpha}} ]$ 
for the even- and odd-channel cases. From eq.~(\ref{eq:Gamma}), we can evaluate
$\langle \Gamma_{\alpha}^{2} \rangle$ and $\langle \Gamma_{\alpha} \rangle^{2}$.
Substituting these quantities into eq.~(\ref{eq:var1}), we obtain
\begin{align}
\mathrm{Var}
\left[
g_{\mathrm{odd}} 
\right]
&= \mathrm{Var} 
\left[
g_{\mathrm{even}} 
\right]
\nonumber \\
&= \frac{2}{15(1+s)^{6}} (s^{6} +6s^{5} +15s^{4} +20s^{3} +15s^{2} )
\nonumber \\
&\ \ \ \ \ \ \ +
m^{-1} \frac{1}{315(1+s)^{8} }
(4s^9 +36s^8 +144s^7 +336s^6 
\nonumber \\
&\ \ \ \ \ \ \ +504s^5 +525s^4 +1050s^3 +315s^2) 
+O \left( m^{-2} \right) .\label{eq:var}
\end{align}
The variance does not depend on whether the total number of channels is even or odd.

\subsection{Long-wire regime} 
We adapt Pichard's treatment~\cite{Pich} to our problem to study the
behavior of $g_{\alpha}$ in the long-wire regime.
It is convenient to
introduce a new set of variables $x_{a}$, related to $\lambda_{a}$ by~\cite{Bee} 
\begin{align}  
\lambda_{a}
=\sinh^{2} x_{a}
,\label{eq:lamdax} 
\end{align}
where $x_{a} \geq 0$. In the odd-channel case, 
the perfectly conducting channel is characterized by
$x_{2m+1} =0$. If we make a change of variables from $\lambda_{a}$ to
$x_{a}$, the evolution equation for $P(x_{1} ,...,x_{m} ;s)$ becomes 
\begin{align}
\frac{\partial P}{\partial s}
=
\frac{1}{4(N_{\alpha} -1)}
\sum_{a=1}^{m} \frac{\partial}{\partial x_{a} }
\left(
\frac{\partial P}{\partial x_{a}}
+P\frac{\partial \Omega_{\alpha} }{\partial x_{a} }
\right)
\label{eq:DMPKlong}
\end{align}
with 
\begin{align}
\Omega_{\alpha}
\equiv
-\ln 
\left(
\prod_{a=1}^{m}
\sinh
2x_{a} \cdot
J_{\alpha}
\right)
.\label{eq:omega}
\end{align}
The explicit form of $\Omega_{\alpha}$ is given by
\begin{align}
\Omega_{\alpha}
& \equiv
-\sum_{a=1}^{m}
\ln |\sinh 2x_{a} |
-4\sum_{b=1}^{m-1}
\sum_{a=b+1}^{m}
\ln |\sinh^{2} x_{a} -\sinh^{2} x_{b} |
\nonumber \\
& \ \ \ \ \ \ \ \ \ \ \ \ \ \ \ \ \ \ \ \ \ \ \ 
\ \ \ \ \ \ \ \ \ \ 
-2\sum_{a=1}^{m} \ln |\sinh^{2} x_{a} |
\delta_{\alpha ,\mathrm{odd} }
.\label{eq:omega2}
\end{align}
In the limit of $s/N_{\alpha } =L/N_{\alpha} l\to \infty$,
we expect that the variables $x_{a} (1\leq a \leq m)$
are much larger than unity and are widely separated.
We thus assume that $1\ll x_{1} \ll x_{2} \ll \cdots \ll x_{m}$.
Under this assumption, we can approximate that $\Omega_{\alpha} \sim -2\sum_{a=1}^{m}
(4a-3+2\delta_{\alpha ,\mathrm{odd}})x_{a} +\mathrm{constant}$.
Substituting this into eq.~(\ref{eq:DMPKlong}), we obtain
\begin{align}
\frac{\partial P_{\alpha} }{\partial s}
=
\frac{1}{2\gamma_{\alpha} }
\sum_{a=1}^{m} \frac{\partial}{\partial x_{a} }
\left(
\frac{\partial P_{\alpha} }{\partial x_{a}}
-2(4a-3+2\delta_{\alpha ,\mathrm{odd}})P_{\alpha}
\right)
,\label{eq:DMPKlong2}
\end{align}
where 
\begin{align}
\gamma_{\alpha} =2(N_{\alpha} -1). \label{eq:gamma} 
\end{align}
The solution of the above equation is 
\begin{align}
P_{\alpha} (x_{1} ,...,x_{m};s)
=
\left(
\frac{\gamma_{\alpha} }{2\pi s}
\right)^{\frac{m}{2} }
\prod_{a=1}^{m}
\mathrm{e}^{-\frac{\gamma_{\alpha}}{2s} \left(
x_{a} -\frac{l}{\xi_{\alpha ,a} } s
\right)^2 }
, \label{eq:pevenodd}
\end{align}
where 
\begin{align}
\xi_{\alpha ,a}
=
\frac{\gamma_{\alpha} l}{4a-3+2\delta_{\alpha ,\mathrm{odd}}}
.\label{eq:kiyokuzai}
\end{align}

We observe from eq.~(\ref{eq:pevenodd})
that each eigenvalue obeys a Gaussian and 
that the even-odd difference appears in 
the distribution function.
From the resulting distribution, we find that
the root-variance $\sqrt{\mathrm{Var} [x_{a}] }
\sim \sqrt{s/\gamma_{\alpha} }$
is indeed much smaller than their spacing
\begin{align}
\langle x_{a+1}
-x_{a} \rangle \sim s/\gamma_{\alpha} .\label{eq:kankaku}
\end{align}
Note that $\langle x_{1} \rangle \sim s/\gamma_{\alpha}$.
Thus, when $s/\gamma_{\alpha} \gg 1$, the dimensionless conductance 
$g_{\alpha} \equiv \delta_{\alpha ,\mathrm{odd}} +2\Gamma_{\alpha}$ 
is dominated by $x_{1}$.
That is to say, considering eq.~(\ref{eq:Gammadef}) and relation of 
$T_{a} = 1/\cosh^2 x_{a}$, $\Gamma_{\alpha}$ behaves as 
\begin{align}
\Gamma_{\alpha}
\approx
4\mathrm{e}^{-2x_{1}}
.\label{eq:Gammax1}
\end{align}
Equation (\ref{eq:pevenodd}) gives
the distribution function for $x_{1}$ as
\begin{align}
p_{\alpha} (x_{1} ;s)=  
\left(
\frac{\gamma_{\alpha} }{2\pi s}
\right)^{\frac{1}{2} }
\mathrm{e}^{-\frac{\gamma_{\alpha} }{2s} 
\left(
x_{1}
-\frac{l}{\xi_{\alpha ,1}} s
\right)^{2} }
.\label{eq:pa}
\end{align}
We will discuss this even-odd difference later.
Using the above distribution function,
the ensemble averages 
$\langle x_{1} \rangle$ and $\langle x_{1}^{2} \rangle$ are obtained as
\begin{align}
\langle x_{1} \rangle
&= \frac{s}{\gamma_{\alpha}} (1+2\delta_{\alpha ,\mathrm{odd}}) 
,\label{eq:x1} \\
\langle x_{1}^{2} \rangle
&=
\frac{s}{\gamma_{\alpha } } +\frac{(1+8\delta_{\alpha ,\mathrm{odd}} )s^2}{\gamma_{\alpha}^{2} }
,\label{eq:xa2}
\end{align}
Based on eq.~(\ref{eq:Gammax1}), we evaluate the ensemble-average
$-\langle \ln \Gamma_{\alpha} \rangle =2\langle x_{1}  \rangle$.
From eqs.~(\ref{eq:gamma}) and (\ref{eq:x1}), we obtain 
\begin{align}
-\langle \ln \Gamma_{\mathrm{odd}} \rangle
&= 
\frac{3L}{2ml} ,\label{eq:x1odd} \\
-\langle \ln \Gamma_{\mathrm{even}} \rangle
&= 
\frac{L}{(2m-1)l} .\label{eq:x1even} 
\end{align}
We estimate the decay length $\xi_{\alpha}$ of the conductance
by identifying 
\begin{align}
\exp[\langle \ln \Gamma_{\alpha} \rangle ] \equiv 
\exp[-2L/\xi_{\alpha}] .
\end{align}
We find that
\begin{align}
\xi_{\mathrm{odd} }
&=
\frac{2}{3}
(N_{\rm odd} -1)l
=
\frac{4}{3} ml, \label{eq:xiodd} \\
\xi_{\mathrm{even} }
&=
2(N_{\rm even} -1)l
=
2(2m-1)l. \label{eq:xieven}
\end{align}
Note that $\xi_{\mathrm{odd}} <\xi_{\mathrm{even}}$ for any positive integer $m$.
This indicates that $g_{\mathrm{odd}}$ decays much faster than $g_{\mathrm{even}}$. 
From eqs.~(\ref{eq:Gammax1}) and (\ref{eq:xa2})-(\ref{eq:x1even}),
we also obtain 
\begin{align}
\sqrt{
\mathrm{Var}
[
\ln \Gamma_{\mathrm{odd} }
]
}
&=
\sqrt{\frac{L}{ml} }
,\label{eq:varlogodd} \\
\sqrt{
\mathrm{Var}
[
\ln \Gamma_{\mathrm{even} }
]
}
&=
\sqrt{\frac{2L}{(2m-1)l} }
.\label{eq:varlogeven}
\end{align}
Note that 
$
\sqrt{
\mathrm{Var}
[
\ln \Gamma_{\alpha}
]
}
<< |\langle \ln \Gamma_{\alpha} \rangle |
$
for the even- and odd-channel cases.
Next, we obtain the ensemble average $\langle \Gamma_{\alpha}
\rangle$ as
\begin{align} 
\langle \Gamma_{\alpha}
\rangle
&\propto
\int_{0}^{\infty}
dx_{1} 
4\mathrm{e}^{-2x_{1} }
\cdot 
\exp 
\left[
-\frac{\gamma_{\alpha} }{2s}
\left(
x_{1}
-\frac{(1+2\delta_{\alpha ,\mathrm{odd} })s}{\gamma_{\alpha}}
\right)^{2}
\right]   \nonumber \\
&\propto
\mathrm{e}^{-\frac{2L}{\xi_{\alpha}^{\prime} }}
,\label{eq:Gammaint} 
\end{align}
where $\xi_{\mathrm{odd}}^{\prime}$ and $\xi_{\mathrm{even}}^{\prime}$
are given by
\begin{align}
\xi_{\mathrm{odd}}^{\prime}
&=
(N_{\rm odd} -1)l
=
2ml ,\label{eq:xitodd} \\
\xi_{\mathrm{even}}^{\prime}
&=
8(N_{\rm even} -1)l
=
8(2m-1)l .\label{eq:xiteven}
\end{align}
The relation $\xi_{\mathrm{odd}}^{\prime} <\xi_{\mathrm{even}}^{\prime}$
holds in any positive integer $m$.
The above results indicate that 
\begin{align}
\langle g_{\rm even} \rangle & \sim \mathrm{e}^{-2L/\xi_{\rm even}^{\prime}}
,\label{eq:long-geven} \\
\langle \delta g_{\rm odd} \rangle & \equiv \langle g_{\rm odd}
\rangle -1 \sim \mathrm{e}^{-2L/\xi_{\rm odd}^{\prime}}
.\label{eq:long-godd}
\end{align}
We also obtain the variance of $\Gamma_{\alpha}$ as 
$\mathrm{Var} [\Gamma_{\rm even} ] \sim \mathrm{e}^{-2L/\xi_{\rm even}^{\prime}}$
and               
$\mathrm{Var} [\Gamma_{\rm odd} ] \sim \mathrm{e}^{-9L/4\xi_{\rm odd}^{\prime} }$.
This results in
\begin{align}
\mathrm{Var} [g_{\rm even} ] & \sim \mathrm{e}^{-2L/\xi_{\rm even}^{\prime}}            
,\label{eq: vargeven} \\
\mathrm{Var} [g_{\rm odd} ] & \sim \mathrm{e}^{-9L/4\xi_{\rm odd}^{\prime} }.
\label{eq:vargodd}
\end{align}
It should be noted that in the limit of $L/\xi_{\alpha}^{\prime} \to \infty$,
$\langle \Gamma_{\mathrm{odd}} \rangle$ and $\langle \Gamma_{\mathrm{even}} \rangle$
are much smaller than $\sqrt{\mathrm{Var} [\Gamma_{\mathrm{odd}}]} \sim \mathrm{e}^{-9L/8\xi_{\mathrm{odd}}^{\prime} }$
and $\sqrt{\mathrm{Var} [\Gamma_{\mathrm{even}}]} \sim \mathrm{e}^{-L/\xi_{\mathrm{even}}^{\prime}}$, respectively.

As noted below eq.~(\ref{eq:DMPK3}), the even-odd differences obtained
above should be attributed to the eigenvalue repulsion from the
perfectly conducting eigenvalue.
The distribution of transmission eigenvalues 
has been given in eq.~(\ref{eq:pevenodd}).
We find that the distribution of 
$\{ x_{1} ,x_{2} ,...,x_{m} \}$ in the odd-channel case
shift toward larger values than those in the even-channel case.
Indeed, we find from eq.~(\ref{eq:x1}) that the peak of 
$p_{\rm odd} (x_{1} ,1)$ is at $\langle x_{1} \rangle =3s/\gamma_{\rm
odd}$, while that of $p_{\rm even} (x_{1} ,1)$ is at 
$\langle x_{1} \rangle =s/\gamma_{\rm even}$, 
where $\gamma_{\rm odd} \approx \gamma_{\rm even}$ for large $m$.
Since $x_{a}$ and $T_{a}$ are related by $T_{a} =1/\cosh^{2} x_{a}$,
the transmission eigenvalues $\{ T_{1} ,T_{2} ,...,T_{m} \}$  
are reduced in the odd-channel case compared with the even-channel case.
This results in the suppression of $\Gamma_{\rm odd}$. 
This tendency should be interpreted as a manifestation of the
repulsion from the perfectly conducting eigenvalue.

\section{Comparison with the SSF Approach}
\begin{figure}[htb]
\begin{center}
\includegraphics[height=6cm]{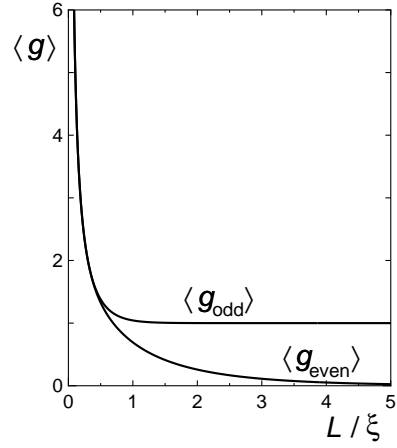}
\end{center}
\caption{The averaged dimensionless conductance in the odd-channel case,
$\langle g_{\rm odd} \rangle$, and that in the even-channel case,
$\langle g_{\rm even} \rangle$, as a function of $L/\xi$.
}
\label{g_ssflin}
\end{figure}

\begin{figure}[htb]
\begin{center}
\includegraphics[height=6cm]{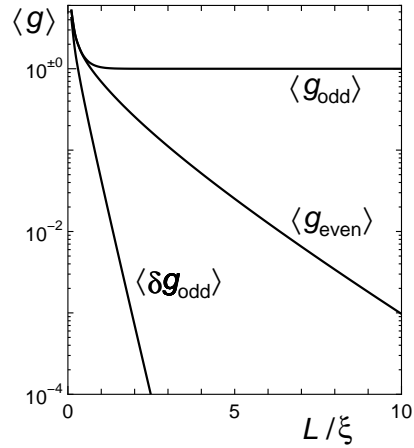}
\end{center}
\caption{$\langle g_{\rm odd} \rangle$, $\langle g_{\rm even} \rangle$ and
$\langle \delta g_{\rm odd} \rangle \equiv \langle g_{\rm odd} \rangle - 1$
as a function of $L/\xi$.
}
\label{g_ssflog}
\end{figure}

In this section, we compare our results for $\langle g_{\alpha} \rangle$ 
and ${\rm Var} [g_{\alpha} ]$ ($\alpha =$ even or odd) given in the previous
section with those obtained from the supersymmetric field (SSF) approach.
We first present the analytic expressions for
$\langle g_{\alpha} \rangle$ and $\langle g_{\alpha}^{2} \rangle$
using the Zirnbauer's super-Fourier analysis.~\cite{Super1,Super2,YT2}
These expressions have been derived previously
except for $\langle g_{\rm odd}^{2} \rangle$,
so we describe their derivation very briefly.
The super-Fourier analysis is based on the nonlinear $\sigma$ model derived
by the SSF approach~\cite{altland} and is valid for
the thick-wire limit which is defined by $N \to \infty$, $L/l \to \infty$
with a fixed $Nl/L$, where $N$ and $l$ are the number of conducting
channels and the mean free path, respectively.
The conductance and its moments are expressed by the propagator,
called the heat kernel, for the nonlinear $\sigma$ model.
In the super-Fourier analysis, we expand the heat kernel in terms of
the eigenfunctions of the Laplacian defined in the $\sigma$-model space.
Note that in addition to normal modes, the eigenfunctions contain
a zero mode and its subsidiary series.~\cite{Super1}
Using the resulting expansion, Zirnbauer obtained the expression
of the average conductance,
\begin{align}
         \label{eq:av-g-Zirnbauer}
  \langle g \rangle
 & =  \frac{1}{2}T({\rm i},1,1)
    + \sum_{n=3,5,7,\cdots}
      \frac{1}{2} \left\{ T({\rm i},n,n-2) + T({\rm i},n-2,n) \right\}
           \nonumber \\
 & \hspace{5mm}
    + 2^{4} \sum_{n_{1},n_{2}=1,3,5,\cdots}
      \int_{0}^{\infty}{\rm d}\lambda \lambda (\lambda^{2}+1)
      \tanh \left( \frac{\pi \lambda}{2} \right)
      n_{1}n_{2} p_{1}(\lambda, n_{1}, n_{2})
           \nonumber \\
 &  \hspace{5mm} \times
      \prod_{\sigma,\sigma_{1},\sigma_{2}=\pm 1}
      (- 1 + {\rm i}\sigma \lambda + \sigma_{1}n_{1} + \sigma_{2}n_{2})^{-1}
      T(\lambda,n_{1},n_{2})
\end{align}
with
\begin{align}
  T(\lambda,n_{1},n_{2})
     & = \exp \left\{ - \left( \lambda^{2}+n_{1}^{2}+n_{2}^{2}-1 \right)
                    \frac{L}{2\xi} \right\} \label{eq:Tnn},
            \\
 p_{1}(\lambda,n_{1},n_{2})
     &  = \lambda^{2} + n_{1}^{2} + n_{2}^{2} - 1 \label{eq:pnn}.
\end{align}
Here, $\xi$ in eq.(\ref{eq:Tnn}) is a typical length scale
characterizing the decay of the conductance.
The first (second) term of eq.~(\ref{eq:av-g-Zirnbauer}) corresponds to
the contribution from the zero mode (the subsidiary modes),
and the third term comes from the normal modes.
Equation~(\ref{eq:av-g-Zirnbauer}) indicates that $\langle g \rangle \to 1/2$
when $L \to \infty$.~\cite{Super1,Super2}
This apparently contradicts the scaling theory, which gives
$\langle g_{\rm even} \rangle \to 0$ and $\langle g_{\rm odd} \rangle \to 1$.
Note that the even-odd difference is not
reflected in the Zirnbauer's argument.
Brouwer and Frahm~\cite{brouwer} introduced a parity operation
and classified the eigenfunctions of the Laplacian
into even-parity modes and odd-parity modes.
The zero mode and the subsidiary modes have odd parity,
so that the first and second terms are classified into the contribution
from the odd-parity modes.
The part of the third term with $n_{1}+n_{2} \equiv 2$ (mod $4$) is the
contribution from the even-parity modes,
while that with $n_{1}+n_{2} \equiv 0$ (mod $4$) is the contribution
from the odd-parity modes.
They pointed out that the odd-parity modes are
unphysical in the even-channel case.
We must neglect the contribution from the odd-parity modes in calculating
$\langle g_{\rm even} \rangle$, and double the contribution from
the even-parity modes to take Kramers degeneracy into account.
We obtain~\cite{brouwer}
\begin{align}
      \label{eq:av-g-even}
  \langle g_{\rm even} \rangle
 & =2^{5} \sum_{\scriptstyle n_{1},n_{2}=1,3,5,\cdots
                  \atop \scriptstyle n_{1}+n_{2}\equiv 2 \, ({\rm mod \, 4})}
      \int_{0}^{\infty}{\rm d}\lambda \lambda (\lambda^{2}+1)
      \tanh \left( \frac{\pi \lambda}{2} \right)
      n_{1}n_{2} p_{1}(\lambda,n_{1},n_{2})
           \nonumber \\
 &  \hspace{5mm} \times
      \prod_{\sigma,\sigma_{1},\sigma_{2}=\pm 1}
      (- 1 + {\rm i}\sigma \lambda + \sigma_{1}n_{1} + \sigma_{2}n_{2})^{-1}
      T(\lambda,n_{1},n_{2}) .
\end{align}
Although Brouwer and Frahm correctly pointed out that the odd-parity
contribution must be excluded in the even-channel case,
they were not aware that the odd-parity modes become essential
in the odd-channel case.
One of the present authors~\cite{YT2} pointed out that
we must exclude the even-parity modes in the odd-channel case.
Neglecting the contribution from the even-parity modes
and doubling the contribution from the odd-parity modes
to take Kramers degeneracy into account,
we obtain~\cite{YT2}
\begin{align}
      \label{eq:av-g-odd}
  \langle g_{\rm odd} \rangle
 & =  T({\rm i},1,1)
    + \sum_{n=3,5,7,\cdots}
         \left\{ T({\rm i},n,n-2) + T({\rm i},n-2,n) \right\}
           \nonumber \\
 & \hspace{5mm}
    + 2^{5} \sum_{\scriptstyle n_{1},n_{2}=1,3,5,\cdots
                  \atop \scriptstyle n_{1}+n_{2}\equiv 0 \, ({\rm mod \, 4})}
      \int_{0}^{\infty}{\rm d}\lambda \lambda (\lambda^{2}+1)
      \tanh \left( \frac{\pi \lambda}{2} \right)
      n_{1}n_{2} p_{1}(\lambda,n_{1},n_{2})
           \nonumber \\
 &  \hspace{5mm} \times
      \prod_{\sigma,\sigma_{1},\sigma_{2}=\pm 1}
      (- 1 + {\rm i}\sigma \lambda + \sigma_{1}n_{1} + \sigma_{2}n_{2})^{-1}
      T(\lambda,n_{1},n_{2}) .
\end{align}
We observe that $\langle g \rangle =(\langle g_{\rm even} \rangle + \langle g_{\rm odd} \rangle)/2$.
On the basis of eqs.~(\ref{eq:av-g-even}) and (\ref{eq:av-g-odd}),
we numerically calculate $\langle g_{\rm even} \rangle$ and
$\langle g_{\rm odd} \rangle$.
The results are displayed in Figs.~\ref{g_ssflin} and \ref{g_ssflog}.
We observe that $\langle g_{\rm odd} \rangle = \langle g_{\rm even} \rangle$
when $L/\xi \ll 1$
and that $\langle g_{\rm odd} \rangle \to 1$
and $\langle g_{\rm even} \rangle \to 0$ when $L/\xi \gg 1$.
The behavior of $\langle g_{\alpha} \rangle$ is in agreement with our
scaling theory. 
Furthermore, we observe from eqs.(\ref{eq:av-g-even}) and (\ref{eq:av-g-odd})
that $\langle g_{\rm even} \rangle \sim {\rm e}^{-L/2\xi}$
and $\langle \delta g_{\rm odd} \rangle \equiv \langle g_{\rm odd} \rangle -1 \sim \mathrm{e}^{-4L/\xi}$.
That is, the characteristic length scale for $\langle \delta g_{\rm odd} \rangle$
is a factor $8$ shorter than that for $\langle g_{\rm even} \rangle$.
This result is also in agreement with the scaling theory which yields 
$\xi_{\rm even}^{\prime} /\xi_{\rm odd}^{\prime} \to 8$ for large $m$.
Let us next consider the variance defined by ${\rm Var} [g_{\alpha} ]\equiv
\langle g^{2}_{\alpha} \rangle - \langle g_{\alpha} \rangle^{2}$.
We need the second moment $\langle g^{2}_{\alpha} \rangle$
to obtain ${\rm Var} [g_{\alpha} ]$.
Zirnbauer and co-workers~\cite{Super2} found
the expression for the second moment,
\begin{align}
  \langle g^{2} \rangle
 & =  \frac{1}{2}T({\rm i},1,1)
    + \sum_{n=3,5,7,\cdots}
      \frac{1}{2} \left( \frac{1}{4}(n-1)^{2}+1 \right)
        \left\{ T({\rm i},n,n-2) + T({\rm i},n-2,n) \right\}
           \nonumber \\
 & \hspace{5mm}
    + 2^{4} \sum_{n_{1},n_{2}=1,3,5,\cdots}
      \int_{0}^{\infty}{\rm d}\lambda \lambda (\lambda^{2}+1)
      \tanh \left( \frac{\pi \lambda}{2} \right)
      n_{1}n_{2} p_{2}(\lambda,n_{1},n_{2})
           \nonumber \\
 &  \hspace{5mm} \times
      \prod_{\sigma,\sigma_{1},\sigma_{2}=\pm 1}
      (- 1 + {\rm i}\sigma \lambda + \sigma_{1}n_{1} + \sigma_{2}n_{2})^{-1}
      T(\lambda,n_{1},n_{2}) \label{eq:g^2-0}
\end{align}
with
\begin{equation}
  p_{2}(\lambda, n_{1}, n_{2}) = \frac{1}{4}
           \left\{  n_{1}^{4} + n_{2}^{4}
                 + (3\lambda^{2} + 1)(n_{1}^{2} + n_{2}^{2})
                 + 2\lambda^{4} - 2\lambda^{2} - 2 \right\} \label{eq:p2nn}.
\end{equation}
Again, the correct result for the even-channel case is obtained if we neglect
the odd-parity contribution and double the even-parity contribution.
The result is~\cite{brouwer}
\begin{align}
       \label{eq:g^2-even}
  \langle g^{2}_{\rm even} \rangle
 & = 2^{5} \sum_{\scriptstyle n_{1},n_{2}=1,3,5,\cdots
                  \atop \scriptstyle n_{1}+n_{2}\equiv 2 \, ({\rm mod \, 4})}
      \int_{0}^{\infty}{\rm d}\lambda \lambda (\lambda^{2}+1)
      \tanh \left( \frac{\pi \lambda}{2} \right)
      n_{1}n_{2} p_{2}(\lambda,n_{1},n_{2})
           \nonumber \\
 &  \hspace{5mm} \times
      \prod_{\sigma,\sigma_{1},\sigma_{2}=\pm 1}
      (- 1 + {\rm i}\sigma \lambda + \sigma_{1}n_{1} + \sigma_{2}n_{2})^{-1}
      T(\lambda,n_{1},n_{2}) .
\end{align}
Similar to the derivation of $\langle g^{2}_{\rm even} \rangle$,
we obtain
\begin{align}
        \label{eq:g^2-odd}
  \langle g^{2}_{\rm odd} \rangle
 & =  T({\rm i},1,1)
    + \sum_{n=3,5,7,\cdots}
      \left( \frac{1}{4}(n-1)^{2}+1) \right)
        \left\{ T({\rm i},n,n-2) + T({\rm i},n-2,n) \right\}
           \nonumber \\
 & \hspace{5mm}
    + 2^{5} \sum_{\scriptstyle n_{1},n_{2}=1,3,5,\cdots
                  \atop \scriptstyle n_{1}+n_{2}\equiv 0 \, ({\rm mod \, 4})}
      \int_{0}^{\infty}{\rm d}\lambda \lambda (\lambda^{2}+1)
      \tanh \left( \frac{\pi \lambda}{2} \right)
      n_{1}n_{2} p_{2}(\lambda,n_{1},n_{2})
           \nonumber \\
 &  \hspace{5mm} \times
      \prod_{\sigma,\sigma_{1},\sigma_{2}=\pm 1}
      (- 1 + {\rm i}\sigma \lambda + \sigma_{1}n_{1} + \sigma_{2}n_{2})^{-1}
      T(\lambda,n_{1},n_{2}) .
\end{align}
\begin{figure}[htb]
\begin{center}
\includegraphics[height=6cm]{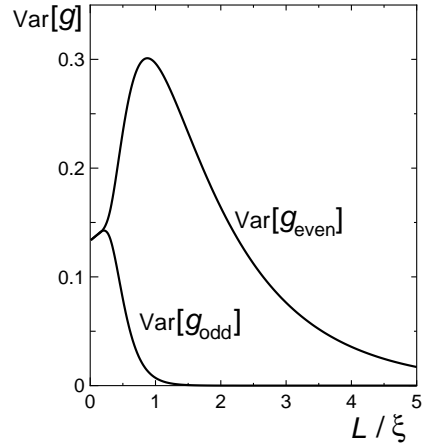}
\end{center}
\caption{The variance of the dimensionless conductance
in the odd-channel case, ${\rm Var} [g_{\rm odd} ]$,
and that in the even-channel case, ${\rm Var} [g_{\rm even}]$,
as a function of $L/\xi$.
}
\label{g_var_lin}
\end{figure}
\begin{figure}[htb]
\begin{center}
\includegraphics[height=6cm]{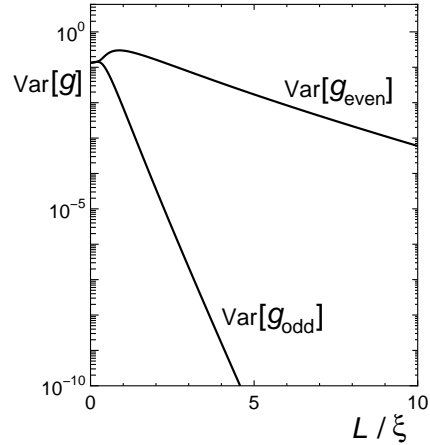}
\end{center}
\caption{${\rm Var} [g_{\rm odd} ]$ and ${\rm Var} [g_{\rm even} ]$
as a function of $L/\xi$.
}
\label{g_varlog}
\end{figure}
On the basis of eqs.~(\ref{eq:av-g-even}), (\ref{eq:av-g-odd}),
(\ref{eq:g^2-even}) and (\ref{eq:g^2-odd}),
we numerically calculate 
${\rm Var} [g_{\rm even} ]$ and ${\rm Var} [g_{\rm odd} ]$.
The results are displayed in Figs.~\ref{g_var_lin} and \ref{g_varlog}.
We observe that ${\rm Var} [g_{\rm odd} ] = {\rm Var} [g_{\rm even} ]$
when $L$ is much shorter than $\xi$.
This is in agreement with our analysis for the short-wire regime.
With increasing $L/\xi$,
${\rm Var} [g_{\rm odd} ]$ very rapidly decreases to zero,
while the decrease of ${\rm Var} [g_{\rm even} ]$ is much slow.
Furthermore, we observe from 
eqs.~(\ref{eq:av-g-even}), (\ref{eq:av-g-odd}), (\ref{eq:g^2-even}) and (\ref{eq:g^2-odd}) that
${\rm Var} [g_{\rm even} ] \sim {\rm e}^{-L/2\xi}$ and
${\rm Var} [g_{\rm odd} ]\sim {\rm e}^{-9L/2\xi}$.
Again, these results are in agreement with the scaling theory.

\section{Conclusion} 
We have studied the quantum electron transport
in disordered wires with symplectic symmetry.
In the symplectic class, 
the reflection matrix $r$
has the scattering symmetry ${}^tr =-r$, from which 
the presence of a perfectly conducting channel without backscattering 
can be proved for the odd-channel case.
The perfectly conducting channel is present even in the long-wire limit.
This indicates the absence of Anderson localization in the odd-channel case.
Since the odd-channel case is very different from the ordinary
even-channel case, we separately consider the two cases.
To study the two cases within a unified manner, we adopt the scaling approach based on a random-matrix theory.
The transport property is fully described by the 
probability distribution function of transmission eigenvalues.
For both the even- and odd-channel cases, we have derived the DMPK
equation which describes the evolution of the distribution function with
increasing system length.
In the even-channel case, the resulting DMPK equation 
is equivalent to the previous result for the symplectic class.
The DMPK equation for the odd-channel case
is different from that for the even-channel case
reflecting the presence of a perfectly conducting channel. 
Using the resulting DMPK equations for the
even- and odd-channel cases, we study the behavior of the dimensionless conductance
for the short-wire regime of $L/\xi_{\alpha} \ll 1$
and the long-wire regime of $L/\xi_{\alpha} \gg 1$,
where $L$ is wire length and $\xi_{\alpha}$ is the conductance decay
length ($\alpha =\mathrm{even} \ \mathrm{or} \ \mathrm{odd}$).

In the short-wire regime, we have derived the
evolution equations for various quantities, such as 
$\langle \Gamma_{\alpha}^{p} \rangle$ and 
$\langle \Gamma_{\alpha}^{p} \Gamma_{\alpha 2} \rangle$,
from which we obtain 
$\langle g_{\alpha} \rangle$ and $\mathrm{Var} [g_{\mathrm{\alpha}} ]$.
We have solved the resulting equations by using the series expansion 
with respect to $s/m$.~\cite{MaS}
We observed that the weak-antilocalization correction to 
$\langle g_{\mathrm{odd}} \rangle$ is equivalent to that to $\langle g_{\mathrm{even}} \rangle$.
Furthermore, we obtained higher order terms of the
series expansion with respect to $s/m$ and found a small even-odd difference
between $\langle g_{\mathrm{odd}} \rangle$ and $\langle g_{\mathrm{even}} \rangle$.
This means that $\langle g_{\mathrm{odd}} \rangle \sim \langle g_{\mathrm{even}} \rangle$.
We have discussed this behavior from the viewpoint of the eigenvalue
repulsion from the perfectly conducting channel.
We have also shown that 
$\mathrm{Var} [g_{\mathrm{odd}} ] = \mathrm{Var} [g_{\mathrm{even}} ]$.
We thus concluded that clear even-odd differences do not appear in the
transport properties in the short-wire regime. 
In the long-wire regime, we approximately solved the DMPK equation 
and studied transport property for the even- and
odd-channel cases.
We have shown that the dimensionless conductance 
in the even-channel case decays exponentially as 
$\langle g_{\mathrm{even}} \rangle \to 0$
with increasing system length,
while $\langle g_{\mathrm{odd}} \rangle \to 1$
in the odd-channel case.
We evaluated the conductance decay length $\xi_{\alpha}$
by identifying $\exp[\langle \ln \Gamma_{\alpha} \rangle] \equiv \exp[-2L/\xi_{\alpha}]$, 
where $\Gamma_{\alpha} =(g_{\alpha} -\delta_{\alpha ,\mathrm{odd}} )/2$,
and found $\xi_{\mathrm{odd}} <\xi_{\mathrm{even}}$.
That is, $g_{\mathrm{odd}} -1$ decays much faster than $g_{\mathrm{even}}$.
We have discussed this behavior from the viewpoint of the eigenvalue repulsion.
We also obtained variance $\sqrt{ \mathrm{Var} [\ln \Gamma_{\alpha} ]}$ 
and observed a clear even-odd difference.

From what has been discussed above,
we conclude that the perfectly conducting channel induces the clear
even-odd difference in the long-wire regime.

\section{Acknowledgement}
This work was supported in part by a Grant-in-Aid for Scientific
Research (C) from the Japan Society for the Promotion of Science.

\appendix
\section{Solution of the coupled equations}
We substitute eqs.~(\ref{eq:fseri})-(\ref{eq:yseri}) into eqs.~(\ref{eq:Tfunc})-(\ref{eq:TT23}) and equate the coefficients of
various powers of $m$. Accordingly, we obtain the following
differential equations, 
\begin{align}
&\ f_{\alpha ,p,0}^{\prime } (s) + pf_{\alpha ,p+1,0} (s) =0 ,\label{eq:f0equ} \\
&\ j_{\alpha ,p,0}^{\prime } (s)
+(p+3)j_{\alpha ,p+1,0} (s) =2f_{\alpha ,p+1,0} (s) ,\label{eq:j0equ} \\
&\ f_{\alpha ,p,1}^{\prime } (s) + pf_{\alpha ,p+1,1} (s) =\frac{1}{2} 
\left[
pj_{\alpha ,p,0} (s) + f_{\alpha ,p,0}^{\prime } (s) \delta_{\alpha ,\mathrm{even} } -2p
f_{\alpha ,p,0} (s) \delta_{\alpha , \mathrm{odd} }
\right] , \label{eq:f1equ} \\
&\ l_{\alpha ,p,0}^{\prime} (s) 
+(p+6)l_{\alpha ,p+1,0} (s)
= 4j_{\alpha ,p+1,0} (s) ,\label{eq:l0equ} \\
&\ k_{\alpha ,p,0}^{\prime} (s)
+(p+5)k_{\alpha ,p+1,0} (s) 
=6j_{\alpha ,p+1,0} (s) -3l_{\alpha ,p+1,0} (s) ,\label{eq:k0equ} \\
&\ j_{\alpha ,p,1}^{\prime} (s)
+(p+3)j_{\alpha ,p+1,1} (s)
=2f_{\alpha ,p+1,1} (s) +2k_{\alpha ,p,0} (s) 
+\frac{1}{2} (p-1)l_{\alpha ,0} (s)
\nonumber \\
&\ \ \ \ \ \ \ \ \ \ \ \ \ \ \ \ \ \ \ \ \ \ \ \ \ \ \ \ \ \ \ \ \ \ 
+\frac{1}{2} j_{\alpha ,p,0}^{\prime} (s) \delta_{\alpha ,\mathrm{even} }
-\{
1+(p+1)\delta_{\alpha ,\mathrm{odd} }
\} j_{\alpha ,p,0} (s) ,\label{eq:j1equ} \\
&\ f_{\alpha ,p,2}^{\prime} (s) +pf_{\alpha ,p+1,2} (s)
=\frac{1}{2} pj_{\alpha ,p,1} (s) +\frac{1}{2} p(p-1)j_{\alpha ,p-1,0} (s) 
\nonumber \\
&\ \ \ \ \ \ \ \ \ \ \ \ \ \ \ \ \ \ \ \ \ \ \ \ \ \ \ \ \ \ \ \ \ \ 
-\frac{1}{2} p(p-1)k_{\alpha ,p-1,0} (s)
+\frac{1}{2} f_{\alpha ,p,1}^{\prime} (s) \delta_{\alpha ,\mathrm{even} }
\nonumber \\
&\ \ \ \ \ \ \ \ \ \ \ \ \ \ \ \ \ \ \ \ \ \ \ \ \ \ \ \ \ \ \ \ \ \ 
-pf_{\alpha ,p,1} (s) \delta_{\alpha ,\mathrm{odd} } .\label{eq:f2equ} 
\end{align}
Note that the above equations are closed for 
$f_{\alpha ,p,0}$, $j_{\alpha ,p,0}$, $f_{\alpha
,p,1}$, $l_{\alpha ,p,0}$, $k_{\alpha ,p,0}$, $j_{\alpha ,p,1}$
and $f_{\alpha ,p,2}$.
To obtain $f_{\alpha ,p,3} (s)$, we need equations for
$y_{\alpha ,p,0}$, $u_{\alpha ,p,0}$, $t_{\alpha ,p,0}$, 
$l_{\alpha ,p,1}$, $k_{\alpha ,p,1}$, $j_{\alpha ,p,2}$ and $f_{\alpha ,p,3} (s)$
in addition to eqs.~(\ref{eq:f0equ})-(\ref{eq:f2equ}).
By considering the series expansion of $f_{\alpha ,p,0} (s)$
at $s=0$ on the basis of eq.~(\ref{eq:f0equ}), we expect that 
\begin{align} 
f_{\alpha ,p,0} (s)=\frac{1}{(1+s)^{p}} . \label{eq:f0}
\end{align}
We can easily show that eq.~(\ref{eq:f0}) satisfies eq.~(\ref{eq:f0equ})
for arbitrary $s$.
Since the resulting power series for the other coefficients 
$j_{\alpha ,p,0} (s), \ f_{\alpha ,p,1} (s),\cdots$
at $s=0$ are not easily identifiable, it is difficult to derive 
all the coefficients in the above manner.

From the argument by Mello and Stone, a solution of coupled differential equations,
eqs.~(\ref{eq:f0equ})-(\ref{eq:f2equ}), is given by the combination of polynomials in $s$ divided by a power of $(1+s)$.
Such a polynomial is written as
\begin{align}
A_{p} (s)
\equiv \frac{1}{(1+s)^{p+n} } (a_{r} s^{r} +a_{r-1} s^{r-1} +\cdots +a_{0} ),
\label{eq:kateiA}     %Ap(s)%
\end{align}
where $n$ and $r$ are integers and $\{a_{r} \}$ are
coefficients which are independent of $p$.
Regardless of integers $n$ and $r$,
$A_{p} (s)$ satisfies the relation of 
\begin{align}
A_{p+1,0} (s)= \frac{1}{1+s} A_{p,0} (s).\label{eq:Arelation}
\end{align}
Noting this observation, we can systematically obtain
the solutions of eqs.~(\ref{eq:f0equ})-(\ref{eq:f2equ}).
Indeed, if $f_{\alpha ,p,0} (s)$ satisfies eq.~(\ref{eq:Arelation}), 
we can rewrite eq.~(\ref{eq:f0equ}) as 
\begin{align}
f_{\alpha ,p,0}^{\prime} (s)
+\frac{p}{1+s} f_{\alpha ,p,0} (s) =0 \label{eq:first order} 
\end{align}
with the initial condition $f_{\alpha ,p,0} (0)=1$.
Its solution is equivalent to eq.~(\ref{eq:f0}).

We assume that $j_{\alpha ,p,0} (s)$ satisfies eq.~(\ref{eq:Arelation}),
and rewrite eq.~(\ref{eq:j0equ}) as 
a first-order linear differential equation. The solution is 
\begin{align}
j_{\alpha ,p,0} (s) 
=\frac{1}{3(1+s)^{p+3}}
(2s^3 +6s^2 +6s+3). \label{eq:j0kai}
\end{align}
Substituting eq.~(\ref{eq:f0}), its derivative 
and eq.(\ref{eq:j0kai}) into eq.~(\ref{eq:f1equ}), we obtain
\begin{align}
f_{\alpha ,p,1}^{\prime } (s) +pf_{\alpha ,p+1,1} (s)
&= \frac{p}{6(1+s)^{p+3} } 
\left[
2(1-3\delta_{\alpha ,\mathrm{odd} } ) s^{3} \right. +3(1
-5 \delta_{\alpha ,\mathrm{odd} } ) s^{2} \nonumber \\
&\ \ \ \ \ \ \left.  -12s\delta_{\alpha ,\mathrm{odd} } -3\delta_{\alpha ,\mathrm{odd} }
\right] .   \label{eq:f1equ2}
\end{align}
We assume that $f_{\alpha ,p,1} (s)$ is expressed as
\begin{align}
f_{\alpha ,p,1} (s) =p\cdot \chi_{\alpha ,p} (s) , \label{eq:fp+1}
\end{align}
where $\chi_{\alpha ,p} (s)$ satisfies eq.~(\ref{eq:Arelation}).
Under the above assumption, we substitute
$f_{\alpha ,p,1}^{\prime } (s)$ and $f_{\alpha ,p+1,1} (s)$ into 
the left-hand side of eq.~(\ref{eq:f1equ}) 
and obtain a first-order linear differential equation for $\chi_{\alpha ,p}(s)$. 
We solve the resulting equation and obtain
\begin{align}
f_{\alpha ,p,1} (s)
=\frac{p}{6(1+s)^{p+2} }
\left(
(1-3\delta_{\alpha ,\mathrm{odd} })s^{3}
-6s^{2} \delta_{\alpha ,\mathrm{odd} } -3s\delta_{\alpha ,\mathrm{odd} }
\right) .
\label{eq:f1kai} %f1(s)
\end{align}
For $j_{\alpha ,p,1} (s)$, 
we obtain the differential equation,
\begin{align}
j_{\alpha ,p,1}^{\prime} (s)
+(p+3)j_{\alpha ,p+1,1} (s)
&=
\frac{1}{90(1+s)^{p+6}}
\Bigl[
p\{
(50-150\delta_{\alpha ,\mathrm{odd}} )s^6 
\ \ \ \ \ \ \ \ \ \ \ \ \ \ \ \ \ 
\ \ \ \ \ \ \ \ \ \ \ \ \ 
\nonumber \\
&\ +(180-780\delta_{\alpha ,\mathrm{odd}} )s^5
+(240-1650\delta_{\alpha ,\mathrm{odd}} )s^4
\nonumber \\
&\ +(150-1830\delta_{\alpha ,\mathrm{odd}} )s^3
+(45-1125\delta_{\alpha ,\mathrm{odd}} )s^2
\nonumber \\
&\ -360\delta_{\alpha ,\mathrm{odd}} s
-45\delta_{\alpha ,\mathrm{odd}}
\}
\nonumber \\
&\ 
+\{
(46-150\delta_{\alpha ,\mathrm{odd}} )s^6
+(186-810\delta_{\alpha ,\mathrm{odd}} )s^5
\nonumber \\
&\ 
+(330-1800\delta_{\alpha ,\mathrm{odd}} )s^4
+(360-2130\delta_{\alpha ,\mathrm{odd}} )s^3
\nonumber \\
&\ 
+(225-1395\delta_{\alpha ,\mathrm{odd}} )s^2
-450\delta_{\alpha ,\mathrm{odd}} s
-45\delta_{\alpha ,\mathrm{odd}}
\}
\Bigr] . \label{eq:j1equation}
\end{align}
We assume that $j_{\alpha ,p,1} (s)$ is expressed as
\begin{align}
j_{\alpha ,p,1} (s)= p\cdot \eta_{\alpha ,p} (s)+\phi_{\alpha ,p} (s)   
,\label{eq:j1p}
\end{align}
where $\eta_{\alpha ,p} (s)$ and $\phi_{\alpha ,p} (s)$ 
satisfy eq.~(\ref{eq:Arelation}).
Under the above assumption, we substitute
$j_{\alpha ,p,1}^{\prime } (s)$ and $j_{\alpha ,p+1,1} (s)$ into 
the left-hand side of eq.~(\ref{eq:j1equation}) 
and obtain first-order linear differential equations for $\eta_{\alpha ,p}(s)$
and $\phi_{\alpha ,p} (s)$. We solve the resulting equations and 
finally obtain $j_{\alpha ,p,1} (s)$. 
Similarly, we assume that 
$f_{\alpha ,p,2} (s)$ is expressed as 
\begin{align}
f_{\alpha ,p,2} (s)= p^2 \cdot \mu_{\alpha ,p} (s)+p \cdot \nu_{\alpha ,p} (s)
,\label{eq:f2p}
\end{align}
where $\mu_{\alpha ,p} (s)$ and $\nu_{\alpha ,p} (s)$ 
satisfy eq.~(\ref{eq:Arelation}).
Repeating the procedure described above, we obtain 
\begin{align}
f_{\alpha ,p,2} (s)
&= \frac{p}{360(1+s)^{p+4}}
\Bigl[
\{ 
11p-9 +(15p-15) \delta_{\alpha ,\mathrm{odd} }
\} s^{6}
\nonumber \\
& \ \ \ \ \ \ \ \ \ \ \ \ \ \ \ 
+
\{
36p-24 
+(120p-120) \delta_{\alpha ,\mathrm{odd} }
\} s^{5} 
\nonumber \\
& \ \ \ \ \ \ \ \ \ \ \ \ \ \ \ 
+
\{
90p-60 +(240p-240) \delta_{\alpha ,\mathrm{odd} }
\} s^{4}
\nonumber \\
& \ \ \ \ \ \ \ \ \ \ \ \ \ \ \ 
+
\{
120p-150+(180p-180)\delta_{\alpha ,\mathrm{odd} }
\} s^{3}
\nonumber \\
& \ \ \ \ \ \ \ \ \ \ \ \ \ \ \ 
+
\{
90p-90+(45p-45)\delta_{\alpha ,\mathrm{odd} }
\} s^{2}
\Bigr]. \label{eq:f2kai} %f2(s)
\end{align}
To derive $f_{\alpha ,p,3} (s)$, we must add the differential equations
for $y_{\alpha ,p,0}$, $u_{\alpha ,p,0}$, $t_{\alpha ,p,0}$, 
$l_{\alpha ,p,1}$, $k_{\alpha ,p,1}$, $j_{\alpha ,p,2}$ and $f_{\alpha ,p,3} (s)$
to eqs.~(\ref{eq:f0equ})-(\ref{eq:f2equ}). We present only the final result:
\begin{align}
f_{\alpha ,p,3} (s)
&=
\frac{p}{45360(1+s)^{p+6} }
\Bigl[
\{
(161p^2 -369p+196)
\nonumber \\
&\ \ \ \ \ \ \ \ \ \ \ \ \ \ \ \ \ \ \ \ \ \ \ \ \ \ \ \ 
-(693p^2 -1953p+1260)\delta_{\alpha ,\mathrm{odd}}
\} s^9
\nonumber \\
&+
\{
(756p^2 -1368p+630)
-(5544p^2 -15624p+10206)\delta_{\alpha ,\mathrm{odd} }
\} s^8
\nonumber \\
&+
\{
(1890p^2 -2448p+1008)
-(19404p^2 -54684p +36288)\delta_{\alpha ,\mathrm{odd} }
\} s^7
\nonumber \\
&+
\{
(2520p^2 -2142p+1260)-(36288p^2 -106218p+73458)\delta_{\alpha ,\mathrm{odd} }
\} s^6
\nonumber \\
&+
\{
(1890p^2 +3402p-4914)-(39690p^2 -120960p+83916)\delta_{\alpha ,\mathrm{odd} }
\} s^5
\nonumber \\
&+
\{
(9450p-24570)-(24570p^2 -75600p+49140)\delta_{\alpha ,\mathrm{odd} }
\} s^4
\nonumber \\
&+
\{
(18900p-24570)-(6615p^2 -19845p+11340)\delta_{\alpha ,\mathrm{odd} }
\} s^3
\nonumber \\
&+
(5670p-5670)s^2
\Bigr] .\label{eq:fp3s}
\end{align}

\end{document}